\documentclass{elsart}
\begin{document}
\begin{frontmatter}
\title{Importance Sampling in Rigid Body Diffusion Monte Carlo}
\author{Alexandra Viel$^1$,  Mehul V. Patel$^{1,2}$, Parhat Niyaz$^1$, and K. B. Whaley$^1$}
\address{Departments of Chemistry$^1$ and Physics$^2$ \\
and Kenneth S. Pitzer Center for Theoretical Chemistry, \\
University of California, Berkeley, CA 94720-1460\\}

\begin{abstract}
{We present an algorithm for rigid body diffusion 
Monte Carlo with importance sampling, which is based on a rigorous short-time expansion of the Green's function for rotational motion in three dimensions.  
We show that this short-time approximation provides correct sampling of the angular degrees of freedom, and provides a general way to incorporate 
importance sampling for all degrees of freedom.
The full importance sampling algorithm significantly improves both calculational efficiency and accuracy of ground state properties, and allows rotational and bending excitations in molecular van der Waals clusters to be studied directly.}
\end{abstract}
\end{frontmatter}

\vspace{2.cm}



\section{Introduction}
\label{intro}
Diffusion Monte Carlo (DMC) methods provide a powerful theoretical 
approach to the 
study of weakly bound clusters of atoms and molecules.   The favorable
polynomial scaling of DMC enables clusters with large numbers
of degrees of freedom to be studied, provided that suitable schemes for
sampling the multi-dimensional wavefunctions exist.  
In the case of van der Waals clusters containing one or more molecules, 
there often exist several different
time scales, deriving on the one hand from the monomer internal degrees of 
freedom, and on the other hand from
the `external' monomer translations and rotations which are associated with the 
van der Waals modes between monomers.  Typically, there is a large time scale
separation between the high frequency internal vibrations of the monomer
constituents, and lower frequency intermonomer van der Waals modes. 
This time scale separation is often invoked to motivate 
approximations in which one effectively freezes the high frequency degrees 
of freedom when studying the role of the low frequency modes.  When applied 
to intramolecular degrees of freedom using Diffusion Monte
Carlo, this approximation leads to
an algorithm first proposed by Buch~\cite{buch92}, commonly referred to as 
`Rigid Body Diffusion Monte Carlo'
(RBDMC), in which all monomers are treated as non-vibrating, rigid molecules
which are free to rotate and translate in the presence of intermolecular 
interactions.

The first implementation of RBDMC for molecular clusters by Buch employed  
a short-time
Green's function for the free rotation of each molecular monomer in its 
instantaneous principal axis frame\cite{buch92}.  
A convenient Gaussian short-time
approximation for such monomer rotational motion was 
assumed in that work.  
A number of applications of RBDMC have since been carried out, 
all of which use a
combination of this angular sampling for monomer rotation in local 
monomer principal axis frames, together with the conventional
sampling of translational motions of monomers in the laboratory
frame\cite{buch92,buch94,niyaz96,buch96,oliveira97,buch97,buch98,kendall98,severson98,buch99,buch99a,iosue99,jakowski00,clary,sorenson97,benoit00}.
The rigid body rotations are carried out using either Euler angles, or
quaternion algebra~\cite{benoit00}.
Considerable use has been made of this
algorithm for water clusters, where the floppy intermolecular degrees of
freedom are difficult to study with other methods for clusters larger than the
water dimer~\cite{buch92,buch94,buch96,buch97,buch98,buch99,buch99a,clary,sorenson97,benoit00}. 

Nearly all of the RBDMC applications to
date employ DMC in its simplest form, namely unbiased Monte Carlo
sampling in which the ground state wavefunction is sampled directly without
using any trial function.  Exceptions to this include a calculation
for HF-Ar$_n$ for which importance sampling of translational 
degrees of freedom was implemented\cite{niyaz96}, and a calculation 
for rotational excitations in clusters of SF$_6$He$_n$\cite{lee99}.
The latter calculation represents the first application of 
the complete importance sampling algorithm for rotations and
translations which is described 
in detail in this work.
Such biased sampling proves useful since unbiased
sampling, while  adequate for ground states of relatively strongly bound or
fairly rigid clusters,  rapidly becomes less efficient for weakly bound 
systems.
Importance sampling also allows for easier computation of quantities other than the 
energy (e.g. structural quantities) that would require descendant weighting\cite{suhm91,liu74} in 
unbiased DMC.  Furthermore, it is also more amenable to study of excited states than are 
unbiased sampling techniques.  

Biased sampling in DMC is generally achieved with the importance sampling
approach first proposed by Kalos and co-workers
for translational motions~\cite{kalos74}. 
The improved efficiency is dramatic for very weakly bound systems such as
the 
clusters of helium and hydrogen, which are dominated by quantum effects. 
Both pure helium
clusters\cite{whaley94} and doped
helium clusters\cite{whaley98} have been studied with this approach.
Excited states can be obtained from fixed node calculations in which there is
no explicit trial function but where nodal constraints are nevertheless
imposed directly in an otherwise unbiased sampling procedure\cite{anderson95}.
More efficient sampling of excited states within fixed node is often possible by 
using explicit trial functional forms.  Exploration of nodal release, nodal
optimization, spectral evolution and other approaches to go beyond the 
fixed node approximation is an active area of 
research\cite{buch99,anderson91,anderson94,anderson95a,anderson99,kwon96,jones97,blume97,blume99,mccoy99,kalos00,sarsa00}.

In this paper we first outline the standard rigid body diffusion Monte
Carlo algorithm with unbiased sampling (Section \ref{unbiased}). 
Our treatment includes 
a quantum mechanical derivation of the short-time Green's
function for free rotation of a general asymmetric top 
(Section \ref{unbiased_theory}).
For spherical tops,
formal solutions for both the imaginary time, diffusive Green's 
function\cite{furry57} and the
real time Green's function\cite{schulman68} were established a long time ago.
However for a general rigid body, {\it i.e.}, an asymmetric top, formal
solutions have been restricted to eigenfunction expansions
\cite{favro60,favro65}, and no rigorous 
derivation of the expected Gaussian form for small angle diffusion
has been presented. 
A formal solution for this is useful since it enables extension for rotation 
in the
presence of external force fields (to which the quantum forces in biased 
diffusion Monte Carlo are formally equivalent) 
to be naturally and rigorously implemented.  
In Section \ref{unbiased_timestep} 
we show how the use of a rigid body diffusion Monte
Carlo algorithm is motivated, with a detailed
discussion of the factors influencing time step choice
in a weakly bound system composed of both atoms and molecules.  
In Section \ref{importance} we then
extend the rotational Green's function formalism 
to enable direct
importance sampling of rotational degrees of freedom. 
Implementation of the full importance sampled
RBDMC algorithm is described in Section \ref{implement}.  
In Section \ref{applications} we provide some basic applications of this
approach which illustrate its use.  First, we demonstrate explicitly with
detailed tests that such short-time
sampling of the angular degrees of freedom will
cover the entire angular configuration space of a rigid body 
(Section \ref{1d3d}). 
Then we 
show with specific examples taken from studies of two weakly bound
molecule-helium systems, i.e., SF$_6$ $^4$He$_{n}$ and HCN $^4$He$_n$,
that full importance sampled RBDMC considerably improves
the efficiency of calculations for ground states   
(Section \ref{structures}), as well as enabling
calculation of Van der Waals  excitations in such systems (Section \ref{nodes}).  

\section{Rigid Body Diffusion Monte Carlo with Unbiased Sampling}
\label{unbiased}
\subsection{Theory}
\label{unbiased_theory}
The diffusion Monte Carlo method provides a stochastic approach to
solution of the Schr\"odinger equation
\begin{equation}
   i\hbar{\partial |\Psi(t)\rangle \over \partial t} = \hat H|\Psi(t)\rangle.
\label{eq:real_t}
\end{equation}
This is accomplished by first transforming the equation to imaginary time, 
$\tau = i{t \over \hbar}$, 
\begin{equation}
   -{\partial |\Psi(\tau)\rangle \over \partial \tau} = \hat H|\Psi( \tau)\rangle,
\label{eq:im_t}
\end{equation}
and then solving the resulting  relaxation equation with the use of a 
short time
approximation to the Green's function for $\hat H$.  The formal solution to
Eq.~(\ref{eq:im_t}) is 
\begin{equation}
   |\Psi(\tau)\rangle = e^{-\hat H\tau }|\Psi(0)\rangle.
\label{eq:formal}
\end{equation}
Expanding the initial wave function in a complete set of 
eigenfunctions of $\hat H$,$\{|\psi_n\rangle \}$:
\begin{equation}
   |\Psi(0)\rangle = \sum_{n}C_n |\psi_n\rangle,
\label{eq:expand1}
\end{equation}
 we obtain
\begin{equation}
   |\Psi(\tau)\rangle = \sum_{n}C_n  e^{-E_n\tau } |\psi_n\rangle.
\label{eq:exapnd2}
\end{equation}
Choosing the energy scale such that $E_0 = 0$ will guarantee that the ground state 
solution is achieved asymptotically:
\begin{equation}
|\Psi(\tau)\rangle \stackrel{\tau \rightarrow \infty }{\longrightarrow} |\psi_0\rangle   
\label{eq:asymp}
\end{equation}
The solution to Eq.~(\ref{eq:formal}) can be expressed in position 
representation by making use of the short-time Green's function:
\begin{eqnarray}
 \Psi(\vec{Q'},\tau + \Delta\tau) & = &  \int \langle\vec{Q'}| e^{-\hat H\Delta \tau }|\vec{Q}\rangle \Psi(\vec{Q},\tau) d\vec{Q} \\
  & = & \int G(\vec{Q} \to \vec{Q'}, \Delta \tau)  \Psi(\vec{Q},\tau) d\vec{Q}.
\label{eq:green}
\end{eqnarray}
where $Q$ refers to the set of coordinates specifying a single body,
and $\vec{Q}$ to the entire set for the  
$N$-body system. In an all-atom calculation, these coordinates will
consist of the 3$N$-dimensional cartesian coordinates of the $N$-body system,
i.e., $\vec{Q} \equiv \vec{R}$.  We shall use the notation
$R$ to refer to the three-dimensional (center of mass) cartesian coordinates 
of a single body, and $\vec{R}$ to the 3$N$-dimensional cartesian 
coordinates of the $N$-body system.
If  $m$ of these bodies are molecules and have to be treated as rigid bodies,
we have $\vec{Q} \equiv \vec{R}, \vec{\Omega}$, where
$\Omega \equiv (\varphi, \vartheta, \chi)$ denotes the Euler angles
specifying the orientation of the principal axis frame (PAF) of a monomer
in the laboratory reference frame, $\vec{\Omega}$ denotes the 
3$m$-dimensional set of angular coordinates for $m$ monomers, and $\vec{R}$
refers to the combined set of 3$m$ cartesian coordinates of the monomer 
centers of mass, and the 3$n$ atomic coordinates for the $n=N-m$ atoms.

Eq.~(\ref{eq:green}) can be propagated in imaginary time $\tau$ to find the
asymptotic solution, Eq.~(\ref{eq:asymp}), given an accurate short-time
Green's function. The propagation is started from some initial guess for the 
wave function, 
$\Psi(\vec{Q},\tau=0)$, which is represented as an ensemble of 
multi-dimensional configuration points or ``walkers'':
\begin{equation}
\Psi(\vec{Q},\tau=0) \cong  \sum_i \delta (\vec{Q}_i - \vec{Q}).
\label{eq:initial}
\end{equation}
Thus the algorithmical issue reduces to finding an
accurate short time Green's function for a given Hamiltonian $\hat H$, together with
an efficient sampling procedure for the multi-dimensional
integral in Eq.~(\ref{eq:green}). 

The simplest diffusion Monte Carlo procedure samples the 
wave function $\Psi(\vec{Q},\tau)$ directly.  This 
unbiased, or non-importance sampled DMC has been extensively used in the
past for applications to both electronic and nuclear structure and
energetics\cite{suhm91,anderson95,reynolds82}.
We now summarize the conventional implementation of RBDMC 
using unbiased DMC,
for Hamiltonians describing the nuclear motion of 
N-body atomic or molecular systems, with 
\begin{eqnarray}
\hat H = \hat T + \hat V - E_R.
\label{eq:H}
\end{eqnarray}
$E_R$ is a reference energy whose significance will become apparent
below.  
We assume for simplicity that the interaction potential is
pair-wise additive, 
\begin{eqnarray}
\hat V = \sum_{i<j} V(r_{ij},\Omega_i,\Omega_j),
\label{eq:V_pair}
\end{eqnarray}
where $r_{ij}$ is the vector between the centers of mass $R_i$ and $R_j$
of monomers $i$ and $j$, respectively.

Most conventional solutions implicitly utilize the second-order, split operator 
decomposition\cite{feitfleck82} of the Green's function:

\begin{equation}
G = e^{-\hat H\Delta \tau} \approx e^{-\frac{\hat V-E_R}{2}\Delta \tau} 
			e^{-\hat T \Delta \tau}
			e^{-\frac{\hat V-E_R}{2}\Delta \tau}
			+ O((\Delta \tau)^2).
\label{eq:split_op}
\end{equation}
For Hamiltonians involving only translational kinetic 
energy, i.e.,
\begin{equation}
\hat T^t = \sum_{i=1}^{N} -D_i\nabla_{i}^2,
\label{eq:T_trans}
\end{equation}
where $D_i = {\hbar ^2 / 2m_{i} }$, the contribution $e^{-T^t \Delta \tau}$ to
the short time Green's function is given by the solution to the diffusion
equation.  In the position representation this is

\begin{eqnarray}
  G^t(\vec{R} \to \vec{R'}, \Delta \tau) &= &
		\left( {1 \over 4\pi \Delta\tau}\right)^{3N/2}\\
& & 
		\prod_i \left( {1 \over D_i} \right)^{3/2} 
	e^{-(\vec{R'} - \vec{R})^2/4D_i\Delta\tau } 
    e^{-\Delta\tau[(V(\vec{R'},\vec{\Omega'}) + V(\vec{R},\vec{\Omega}) )/2 
	-E_R]} \nonumber. 
\label{eq:G_trans}
\end{eqnarray}
Note that this corresponds formally to rigid bodies moving with fixed 
relative orientations specified by the set of Euler angles $\{ \Omega_i \}$.
In general, this would not be a physically useful representation. However for
a single molecule moving in an atomic cluster, e.g., in a rare gas cluster,
this can sometimes provide a good representation, 
when the molecule is sufficiently
heavy that the molecular rotational kinetic energy is small compared to the
rare gas kinetic energy contributions\cite{barnett93}.
The potential term in Eq. (\ref{eq:split_op}) is diagonal in the position
representation: 
\begin{equation}
\langle \vec{Q} | e^{-\frac{\hat{V}-E_{R}}{2}\Delta \tau} | \vec{Q'} \rangle 
= e^{-\frac{V(\vec{Q})-E_{R}}{2}\Delta \tau} \langle \vec{Q} |\vec{Q'} \rangle. 
\label{eq:poten}
\end{equation}
In the position representation, Eq.~(\ref{eq:split_op}) then becomes a product
of non-local single particle diffusive terms from the kinetic energy, and a 
local term from the potential energy.  Depending on its sign, the latter can 
lead to exponential growth or decay of the amplitude 
$\Psi(\vec{Q})$ in Eq.~(\ref{eq:green}). 

For Hamiltonians involving only rotational kinetic energy, we have
for a general rigid body, i.e., an asymmetric top,
\begin{equation}
\hat T^r = \sum_{p,\alpha}\tilde{d}_{p\alpha} \hat{L}_{p\alpha}^2,
\label{eq:T_rot}
\end{equation}
where $\tilde{d}_{p\alpha} = {1 / 2I_{p\alpha} } $. $I_{p\alpha}$ is the 
principal axis moment of inertia for particle $p$ relative to its principal 
axis $\alpha = x, y, z$, and $\hat{L}_{p\alpha}$ is the corresponding component of the 
angular momentum operator in the $p$th PAF:  
\begin{equation}
{\hat L}_{p\alpha} = \frac{\hbar}{i} \frac{\partial}{\partial \phi_{p\alpha}}, \, \alpha = x,y,z.
\label{eq:L}
\end{equation}
Here $\vec{\phi} \equiv (\phi_x, \phi_y$, $\phi_z)$ specifies the angles of 
one-dimensional rotations around the molecular principal axes 
$x, y$, and $z$,
respectively. 
The rotational Green's function for a general rigid body should formally be 
expressed as $G^r(\Omega \rightarrow \Omega', \Delta \tau)$.
However, applying the second-order short-time decomposition
\begin{equation}
e^{-\hat T^r \Delta \tau} = \prod_p e^{-\hat T^r_p \Delta \tau} 
\end{equation}
where
\begin{equation}
e^{-\hat T^r_p \Delta \tau} =  e^{-\tilde{d}_{x}\hat{L}^2_{px}\Delta \tau} 
			   e^{-\tilde{d}_{y}\hat{L}^2_{py}\Delta \tau} 
			   e^{-\tilde{d}_{z}\hat{L}^2_{pz}\Delta \tau}
                           + O(\Delta \tau^2), 
\label{eq:bk_r}
\end{equation}
leads immediately to the following short-time factorization into one-dimensional
rotational Green's functions:
\begin{equation}
G(\vec{\phi} \to \vec{\phi}',\Delta\tau) = 
	\prod_p G^r_p(\vec{\phi} \to \vec{\phi}',\Delta\tau), 
\label{eq:G_asymm1}
\end{equation}
\begin{eqnarray}
G^r_p(\vec{\phi} \to \vec{\phi}',\Delta\tau) = & & 
    <\phi_x'| e^{-\tilde{d}_{px}\hat{L}^2_{px} \Delta\tau} |\phi_x>
    <\phi_y'| e^{-\tilde{d}_{py}\hat{L}^2_{py} \Delta\tau} |\phi_y>  \nonumber\\
 & &
    <\phi_z'| e^{-\tilde{d}_{pz}\hat{L}^2_{pz} \Delta\tau}|\phi_z>.
\label{eq:G_short}
\end{eqnarray}
Each of the one-dimensional rotational Green's functions can be evaluated
by inserting a complete set of one-dimensional angular momentum eigenstates 
of the angular momentum along the corresponding molecular PAF. This is
given by 
\begin{equation}
\sum_m | L^m_{p\alpha} \rangle \langle L^m_{p\alpha} | = \hat 1, 
\label{eq:complete_Lm}
\end{equation}
where
\begin{equation}
\langle \phi_{p\alpha} | L^m_{p\alpha} \rangle = \frac{1}{\sqrt{2 \pi}} e^{+im \phi_{p\alpha}},
\label{eq:angmom_eigenf}
\end{equation}
with eigenvalues 
\begin{equation}
\hat L_{p\alpha} |L^m_{p\alpha}\rangle = m\hbar|L^m_{p\alpha} \rangle, \, \, m = 0, \pm 1, \pm 2, ...
\label{eq:Lm_values}
\end{equation}
Then, e.g., for $\alpha = x$, we readily obtain
\begin{equation}
\langle \phi'_{px} | e^{-\tilde{d}_{px}\hat{L}^2_{px}\Delta\tau } | \phi_{px} \rangle =
	\frac{1}{2\pi} 
	\sum_m e^{- d_{px} m^2 \Delta\tau + i m (\phi_{px} - \phi'_{px}) } 
\label{eq:onediml}
\end{equation}
where we have used the scaled constant
$d_{p\alpha} = {\hbar}^2 \tilde{d}_{p\alpha} \equiv {\hbar}^2 / 2 I_{p\alpha}$
for particle $p$ and principal axis $\alpha$.  We shall see below that in the
short time limit, $\Delta \tau \rightarrow 0$, this constant 
is identical to the rotational diffusion constant for this degree of freedom. 
Note that it is
also equivalent to the molecular rotation constant $B_{p\alpha}$. As discussed
in Section \ref{unbiased_timestep} below, the magnitude of this relative to 
the masses of the $n=N-m$ atoms, of the $m$ molecules, 
together with the strength of the interaction potentials, 
will determine the time step for a given imaginary time propagation. 

It remains to be shown that Eq.~(\ref{eq:onediml}) reduces to the expected Gaussian form in 
the small 
$\Delta\tau$ limit. We do this by showing the reverse, i.e., by expanding the 
Gaussian form
\begin{equation}
g(\phi) = \frac{1}{\sqrt{4\pi d \Delta\tau}} e^{-\phi^2 / 4d \Delta\tau}
\label{eq:gaussian}
\end{equation}
in eigenfunctions of the one-dimensional angular momentum operator
$\hat{L} = -i\hbar \partial /\partial \phi$:
\begin{equation}
g(\phi) = \sum_{n=0}^{\pm \infty} c_n \frac{1}{\sqrt{2\pi}} e^{i n \phi/ \hbar}.
\label{eq:g_expand}
\end{equation}
Hence
\begin{equation}
c_n =\frac{1}{8 \pi^2 d \Delta\tau} \int_{-\pi}^{\pi} d\phi \, e^{-\frac{\phi^2}{4d\Delta\tau}-i n \phi /\hbar}.
\label{eq:c_n}
\end{equation}
For $\Delta\tau << 1$, we can extend the 
limits of this integral to $\pm \infty$, allowing evaluation to yield
\begin{equation}
c_n = \frac{1}{\sqrt{2\pi}} e^{- n^2 d \Delta\tau }.
\label{eq:c_n_eval}
\end{equation}
Therefore we make the identification
\begin{equation}
\sum_{n=0}^{\pm \infty} \frac{1}{2\pi} e^{- n^2 d \Delta\tau + in \phi} 
\stackrel{\Delta\tau \rightarrow 0} {\simeq}
 \frac{1}{\sqrt{4\pi d \Delta\tau}} e^{-\phi^2 / 4d \Delta\tau}.
\label{eq:gauss_expand}
\end{equation}
Replacing $\phi$ in Eq.~(\ref{eq:gauss_expand}) by $(\phi_{px} - \phi'_{px})$ leads to the 
desired Gaussian form of the sum in Eq.~(\ref{eq:onediml}), i.e.,
\begin{equation}
\frac{1}{2\pi} \sum_m e^{- d_{px} m^2 \Delta\tau + i m(\phi_{px} - \phi'_{px}) } 
\stackrel{\Delta\tau \rightarrow 0} {\simeq}
 \frac{1}{\sqrt{4\pi d \Delta\tau}} e^{-(\phi_{px} - \phi'_{px})^2 / 4d_{px} \Delta\tau}.
\label{eq:equate_gauss}
\end{equation} 
This yields the following short-time factorized limiting form for $G^r_i$: 
\begin{equation}
G^r_p(\vec{\phi} \to \vec{\phi}',\Delta\tau) \stackrel{\Delta\tau \rightarrow 0}{\simeq} 
  \frac{1}{\left( 4\pi \Delta\tau\right)^{3/2}} \prod_{\alpha} 
\frac{1}{\left(d_{p\alpha} \right)^{1/2}} 
       e^{-[(\phi_{p\alpha} -\phi'_{p\alpha})^2/d_{p\alpha}]
/4\Delta\tau}.
\label{eq:G_rot}
\end{equation}
The constants $d_{p\alpha}$ are now seen to act as one-dimensional
diffusion constants for 
each of the rotational degrees of freedom in the $p$-th molecular PAF, as noted
above.

We note that Eq.~(\ref{eq:G_rot}) is formally overcomplete in
angle space, giving rise to a total integrated angular volume of $(8\pi)^3$
instead of the requisite $4\pi^2$.  This is a consequence of the short-time
factorization into three one-dimensional rotations, Eq.~(\ref{eq:bk_r}).  
For a spherical top this can be avoided when the Eulerian short-time limit
described below is employed.
A comparison of the two short-time propagators for a spherical top is provided in 
Section~\ref{applications}, where we 
show that the formal violation of the angular configuration volume deriving from the 
product of one-dimensional rotations does not
cause problems if the time step $\Delta \tau$ is sufficiently small.  The
short-time factorization can be improved upon if necessary by making use of 
the split operator decomposition\cite{feitfleck82}, but we have not found this
necessary in any applications to date.

For the special case of a spherical top 
($d_{p\alpha} \equiv d_p, \alpha = x,y,z$), the exact form of the Green'
s function is well known\cite{schulman68}, as is its short-time diffusive limit\cite{furry57}
\begin{equation}
G^r_p(\Omega \rightarrow \Omega', \Delta \tau) = 
\left( \frac{1}{4 \pi d_p \Delta\tau} \right)^{3/2} 
e^{-\zeta^2 /{4 d_p \Delta \tau}},
\label{eq:G_sph}
\end{equation}
where $\zeta$ is the angle of rotation about an arbitrary axis.
This short-time diffusive limit was already derived by Furry in 1957\cite{furry57}.  
It follows directly from taking the small $\zeta$, small
$\Delta \tau$ limit to the imaginary time transcription of the 
infinitesimal propagator for a spherical top which was derived rigorously by 
Schulman (see Eq. (4.9) in Ref.\cite{schulman68}, with $\Gamma \equiv \zeta$,
and $d_p \equiv \hbar^2 /2I$).  
Eq.~(\ref{eq:G_sph}) corresponds to the Eulerian
description of rotation of a rigid body as a rotation through a single angle 
$\zeta$ about an arbitrary axis $\hat{n}_{\zeta}$\cite{goldstein}.  
It can be implemented
in a Monte Carlo sampling procedure by first choosing a randomly oriented 
axis 
$\hat{n}_{\zeta}$ (uniform sampling of azimuthal angle $\phi$ on the
interval $[0, 2\pi)$ and of $\cos(\theta)$ on the interval $[-1,0]$), 
and then sampling the angle of rotation $\zeta$ 
about this axis
from a Gaussian distribution of width $ \sqrt{3\times 2 d_p \Delta \tau} $. 
The origin of the factor of 3 is 
described in Appendix A.  Related Eulerian rotational sampling has been 
employed in classical Monte Carlo rigid body simulations, but with the Gaussian sampling of the rotation angle $\zeta$ replaced by uniform sampling inside a fixed interval $[-\Delta\zeta_0/2,\Delta\zeta_0/2]$\cite{rao79}.
This satisfies the conditions of
reversibility, detailed balance, and accessibility of all orientations, but does not
correspond to the true Green's function, Eq.~(\ref{eq:G_sph}). It will therefore not 
necessarily yield the 
correct energies and expectation values in a quantum Monte Carlo calculation. 

When the Hamiltonian involves both translational and rotational kinetic
energy terms, the Green's functions $G^t$ and $G^r$ can be combined:
\begin{equation}
G(\vec{R},\vec{\Omega} \rightarrow \vec{R}',\vec{\Omega}', \Delta \tau) 
		\approx
	G^t(\vec{R} \rightarrow \vec{R}', \Delta \tau)
	G^r(\vec{\phi} \rightarrow \vec{\phi}', \Delta \tau).
\label{eq:G_full}
\end{equation}
A multi-dimensional configuration (commonly referred to as a 'walker'),
now carries both translational and angular coordinates.
Sampling of the rotational and translational degrees of freedom are 
carried out by independent diffusive moves in these coordinates, as 
described by the factorization in 
Eq.(\ref{eq:G_full}). However, there will be coupling of the 
translations with rotations via the potential energy term 
$V(\vec{R},\vec{\Omega})$ in Eq. (\ref{eq:split_op}).
Details of the stochastic evaluation of 
Eq. (\ref{eq:G_trans}) is described in Section \ref{implement}.
Use of Eq.~(\ref{eq:G_full}) is usually referred to as 
"rigid body diffusion Monte Carlo" (often abbreviated by its acronym RBDMC),
because of the implicit representation of 
the full $N$-body kinetic energy by a sum of rigid body rotational 
contributions together with the sum of the center of mass translational terms.

\subsection{Time step considerations}\label{unbiased_timestep}
The time step parameter used during the Monte Carlo simulation is determined by two factors.
First, as pointed out by Suhm and Watts \cite{suhm91}, when all other factors
are equal, the time step is controlled by the smallest mass. 
Indeed, for light particles, the diffusion coefficient is large and the time step has to be small enough so that the particle does not move too far in a 
single step.
But this is not the only factor.  The effect of the potential can also be
important.
For example, strongly bound particles must be sampled with a smaller 
displacement, and hence with a smaller time step, to avoid the majority of 
Monte Carlo configurational moves ending in destruction of the configuration.
In the general case of a mixture of light and heavy particles, the time step 
has therefore to be chosen not only according to the smallest mass value, 
but also taking the effect of the potential into account.
The presence of heavy, strongly bound particles may require the use of a 
shorter time step. For example, in
the case of the water dimer, Gregory and Clary \cite{gregory94} use time 
steps of 0.5 - 5.0 au for the all-atom calculations.
Those small values are needed because of the presence of high frequency 
internal modes of the monomers.  When
using RBDMC, they were able to increase the time step up to 30 - 60 au,
because the introduction of the rigid body approximation eliminated these 
strongly confined degrees of freedom.
This decrease in computational time step is generally cited as the primary
motivation for an RBDMC calculation.

The situation with regard to time steps is somewhat more complicated
for helium clusters, where a wide range of time scales is operative.
For pure helium clusters, the time steps used vary typically
from 1000 to 10000 au \cite{barnett93,lewerenz97}.
Introduction of a non-rotating, rigid SF$_6$ impurity requires a time step
reduction to 250 - 500 au \cite{barnett93,mcmahon96} even though 
the mass of SF$_6$ is much bigger than that of He. 
This reduction of time step is due to the fact that the helium-SF$_6$ 
interaction potential is considerably stronger than the He-He potential: 
the larger time step will lead to inefficient sampling of the 
He-SF$_6$ relative motion.
When the molecular rotation is included, the situation becomes quite
complicated as one now has the additional intrinsic time scale of the 
molecular rotation. Thus one needs to
compare the corresponding diffusion coefficients $d_p$ and $D_p$ for each
species $p$ included in the simulation, and to assess both the relative roles
of the masses/moments of inertia (measured via $d$ and $D$), {\it and} the 
confining role of the interactions in all degrees of freedom.  
For the systems studied in this paper, namely SF$_6$ and HCN in clusters
of $^4$He,
the defining parameters are given in Table \ref{Table_d}.  
\begin{table}[b]
\caption[]{Diffusion coefficients and potential characteristics for SF$_6$ and HCN in $^4$He clusters.
For SF$_6$, V$^{sad}$ corresponds to an adiabatic barrier for a helium atom to move between sites.  It corresponds to the minimum of the potential along the C$_2$ symmetry axis.
For HCN, V$^{sad}$ is defined as the maximum of $min_r V(r,\theta)$ for $\theta$ varying from 0 to $\pi$.
It corresponds to the minimum of the potential for $\theta = \pi$.
}
\begin{tabular}{cccccc} \hline
Species (X) & $D$ (au) & $d$ (au) & V$^{min}_{He-X}$ in K &V$^{sad}_{He-X}$ in K & $E_o$ in K\\
\hline
He           & $6.80\;10^{-5}$ & -  & -10.95 & -   &  -\\
SF$_6$         & $1.86\; 10^{-6}$& $4.15\; 10^{-7} $ & -83.86 & -63.13 & -37.4 \\
HCN         & $1.01\; 10^{-5}$& $6.73\; 10^{-6} $ & -42.40  & -30.14  & -13.9 \\
\hline
\end{tabular}
\label{Table_d}
\end{table}          
The time steps required to
eliminate any time step bias
for these two systems are of order 20 - 40 au for SF$_6$, 10 - 30 au for HCN) 
in implementations where attempted moves involve all particles 
being moved together.  Somewhat larger time steps can be used when 
attempted moves involve only single particles 
(e.g., $\sim$ 50 au for SF$_6$ in $^4$He$_n$).
The latter is preferred for all except very diffuse systems. 
For molecules interacting with helium, in general all-atom potentials are not
available, and this is a second reason why RBDMC is essential here.

\section{Rigid Body Diffusion Monte Carlo with Importance sampling}
\label{importance}
 
We consider first the conventional form of importance sampling in
cartesian coordinate systems. This starts with the introduction of a distribution
function $\mathcal{F}(\vec{R},\tau)$ which is biased by a trial function 
$\Psi_T(\vec{R})$:
\begin{equation}
  {\mathcal F}(\vec{R}, \tau) = \Psi(\vec{R}, \tau)\Psi_T(\vec{R}).
\label{eq:f_imp}
\end{equation}
Taking its imaginary time derivative and using Eqs.~(\ref{eq:im_t}) and
(\ref{eq:H}) with $\hat T\equiv \hat T^t$ (Eq.~(\ref{eq:T_trans})), leads to
\begin{eqnarray}
\frac{\partial {\mathcal F}(\vec{R},\tau)}{\partial \tau} & = & [D \nabla^2  
			- D \vec{\nabla} \cdot \vec{F} 
		 	- D \vec{F} \cdot \vec{\nabla}
			- E_L(\vec{R}) + E_R ] {\mathcal F}(\vec{R},\tau) \nonumber \\
			& = & -{\tilde H}^t {\mathcal F}(\vec{R},\tau)
\label{eq:f_difeq}
\end{eqnarray}
where $E_L(\vec{R})$ is the ''local energy'', given by
\begin{equation}
E_L(\vec{R}) = \Psi_T(\vec{R})^{-1} \hat H \Psi_T(\vec{R})
\label{eq:e_local}
\end{equation}
and $\vec{F}(\vec{R})$ is referred to as the ''quantum force'', given
\begin{equation}
\vec{F}(\vec{R}) = 2 {\vec \nabla} \ln \Psi_T(\vec{R})
\label{eq:F_Q}
\end{equation}
For simplicity, we have set $D_p = D$ for all $p = 1...N$.
The solution to Eq.~(\ref{eq:f_difeq}) is then
\begin{eqnarray}
{\mathcal F}(\vec{R'},\Delta \tau) = \int {\tilde G}^t(\vec{R}\rightarrow\vec{R'}, \Delta \tau) {\mathcal F}(\vec{R},0) d \vec{R},
\label{eq:f_soln}
\end{eqnarray}
where the translational importance sampling Green's function ${\tilde G}^t$ is 
formally defined by
\begin{equation}
{\tilde G}^t = \Psi_T^{-1}(\vec{R}) G^t(\vec{R} \rightarrow \vec{R'}, \Delta \tau) \Psi_T(\vec{R'}).
\label{eq:G_imp}
\end{equation}
The short-time solution for ${\tilde G}^t$ is well known\cite{reynolds82}. It is usually
derived within a Fokker-Planck formulation\cite{hammond94}.  
We derive here an alternative solution using operators which provides a 
introduction to the subsequent new derivation of the full translational and rotational importance sampled Green's function, $\tilde{G}$.
${\tilde G}^t$ is initially cast in the position representation, 
\begin{equation}
{\tilde G}^t(\vec{R} \rightarrow \vec{R'}, \Delta \tau)  = 
	\langle \vec{R'} | e^{- \Delta \tau \tilde{H}^t} | \vec{R} \rangle.
\end{equation}
The local energy contributions can then be immediately evaluated: 
\begin{eqnarray}
{\tilde G}^t(\vec{R} \rightarrow \vec{R'}, \Delta \tau)  
 	   &=&  e^{-\Delta \tau [(E_L(\vec{R'}) + E_L(\vec{R}))/2 -E_R]} \nonumber \\
& & 	\langle \vec{R'} | 
	e^{\Delta \tau [ D \nabla^2 - 
	D \vec{\nabla} \cdot \vec{F} - D \vec{F} \cdot \vec{\nabla} ]} 
	| \vec{R} \rangle.
\label{eq:G_oper}
\end{eqnarray}
We now replace $\nabla^2$ by the operator $ - \hat{P}^2/{\hbar^2}$ 
($\hat{P} = - i\hbar \vec{\nabla}$) and use the
identity
\begin{equation}
\vec{F} \cdot \vec{\nabla} = \frac {i} {\hbar} \hat{P} \cdot \vec{F} 
				-\vec{\nabla} \cdot \vec{F}
\label{eq:identity}
\end{equation}
(see Appendix B for proof), to obtain
\begin{equation}
\tilde{G}^t (\vec{R} \rightarrow \vec{R'}, \Delta \tau)  = 
 	e^{-\Delta \tau [(E_L(\vec{R'}) + E_L(\vec{R}))/2 -E_R]}
	\langle \vec{R'} | 
	e^{\Delta \tau [ - \frac{1}{\hbar^2} D \hat{P}^2 - 
		 \frac{i}{\hbar} D \hat{P} \cdot \vec{F} ]}
	| \vec{R} \rangle.
\label{eq:G_ident}
\end{equation}
Inserting a complete set of momentum states $\{ |\vec{p}\rangle \}$, with $\vec{p}$ representing 
the eigenvalues of the momentum operator $\hat{P}$, allows the last factor
in Eq.~(\ref{eq:G_ident}) to be rewritten as
\begin{eqnarray}
\int d\vec{p} \, \langle \vec{R'} | \vec{p} \rangle 
	e^{- \frac{1}{\hbar^2}D{p^2}\Delta\tau}
	e^{-\frac{i}{\hbar}D \vec{p}\cdot\vec{F}(\vec{R})\Delta\tau}
	\langle \vec{p} | \vec{R} \rangle
& =&
\left(\frac{1}{4\pi{\hbar}^2}\right)^{3/2} \\
& & \int d\vec{p} \,
	e^{-\frac{1}{\hbar^2} D p^2 + \frac{i}{\hbar} \vec{p}\cdot
		[\vec{R'}-\vec{R}-D\vec{F}(\vec{R})\Delta\tau]},  \nonumber
\label{eq:integ1}
\end{eqnarray}
where we have used 
\begin{equation}
\langle \vec{R} | \vec{p} \rangle = \left( 2\pi\hbar \right)^{-3/2} 
		e^{\frac{i}{\hbar} \vec{p}\cdot\vec{R}}.
\label{eq:p_eignf}
\end{equation}
Substituting $\vec{k} = \vec{p}/\hbar$, and evaluating Eq.~(\ref{eq:integ1}) using the 
standard integral
\begin{equation}
\int d^3k \, e^{-ak^2 +i\vec{x}\cdot\vec{k}} 
	= \left( \frac{\pi}{a} \right)^{3/2}e^{-x^2/4a},
\label{eq:integral}
\end{equation}
leads to the well known result, 
\begin{eqnarray}
  \tilde{G}^t(\vec{R} \to \vec{R'} ,\Delta \tau) & = & 
		\left({1 \over 4\pi D\Delta\tau}\right)^{3N/2} \nonumber \\
& &
    		e^{-\Delta \tau[(E_L(\vec{R'}) + 
		E_L(\vec{R}) )/2 -E_T]} 	
		 e^{ -[\vec{R'} - \vec{R}
   -D \vec{F}(\vec{R})\Delta\tau]^2/4D\Delta\tau }.
\label{eq:Gtilde_trans}
\end{eqnarray}

Now we expand the discussion to include the rotational kinetic energy
$\hat T^r$, and to incorporate an orientation dependent interaction potential,
$V(\vec{R},\vec{\Omega})$.  The biased distribution function becomes
\begin{equation}
{\mathcal F}(\vec{R},\vec{\Omega},\tau) = \Psi(\vec{R},\vec{\Omega},\tau)
				\Psi_T(\vec{R},\vec{\Omega}).
\label{eq:imp_RO}
\end{equation}
Identifying the quantum forces for one-dimensional angular motion around 
each of the principal axes, $\alpha = x,y,z$,  for each monomer $p$,
\begin{eqnarray}
f_{p\alpha} &=& 2 \frac{\partial}{\partial \phi_{p\alpha}} \ln \Psi_T(\vec{R},
\vec{\Omega}),
\label{eq:f_defs}
\end{eqnarray}
and the angular momentum operators $\hat{L}_p$ in the $p$-th PAF, Eq.~(\ref{eq:L}), 
leads to the generalization of Eq.~(\ref{eq:f_imp}) for a general rigid
body Hamiltonian Eq.~(\ref{eq:H}), with $\hat T = \hat T^t + \hat T^r$ and interaction potential
$\hat V(\vec{R},\vec{\Omega})$:
\begin{eqnarray}
\frac{\partial {\mathcal F}(\vec{R},\vec{\Omega},\tau)}{\partial \tau} & = &  
			[D \nabla^2  
			- D \vec{\nabla} \cdot \vec{F} 
		 	- D \vec{F} \cdot \vec{\nabla} \nonumber \\
& & 
+ \sum_{p\alpha} d_{p\alpha} \frac{\partial^2}{\partial \phi_{p\alpha}^2}
- \sum_{p\alpha} \left\{ d_{p\alpha} \frac{\partial}{\partial \phi_{p\alpha}} f_{i\alpha} 
- d_{i\alpha} f_{p\alpha} \frac{\partial}{\partial \phi_{p\alpha}} \right\}
	\nonumber \\
& & - E_L(\vec{R},\vec{\Omega}) + E_R ] {\mathcal F}(\vec{R},\vec{\Omega}\tau) \nonumber \\
			  & = &  -{\tilde H} {\mathcal F}(\vec{R},\vec{\Omega}\tau).
\label{eq:f_full_difeq}
\end{eqnarray}
This can be simplified 
using Eq.~(\ref{eq:identity}) and its analog for angular
motions (Appendix B), 
\begin{equation}
f_{\alpha} \frac{\partial}{\partial \phi_{\alpha}} = \frac {i} {\hbar} 
	\hat{L}_{\alpha} f_{\alpha} - \frac{\partial f_{\alpha}}{\partial \phi_{\alpha}},
\label{eq:identity^r}
\end{equation} 
to arrive at
\begin{eqnarray}
\frac{\partial {\mathcal F}(\vec{R},\vec{\Omega},\tau)}{\partial \tau} & = & 
		\left[- \frac{D}{\hbar^2} \hat{P}^2  
		- i \frac{D}{\hbar} \hat{P} \cdot \vec{F} 
		- \sum_{p\alpha}  \tilde{d}_{p\alpha} {\hat L}^2_{p\alpha}  \right.\\
& & \left.
		-i \hbar \sum_{p\alpha} \tilde{d}_{p\alpha} {\hat L}_{p\alpha} f_{p\alpha}
		- E_L(\vec{R}) + E_R \right] {\mathcal F}(\vec{R},\vec{\Omega}, \tau) 
					\nonumber \\
			& = & -{\tilde H} {\mathcal F}(\vec{R},\vec{\Omega},\tau).
\label{eq:fr_difeq}
\end{eqnarray}

The short-time Green's function for ${\tilde H}$ is put into position
representation using Eqs.~(\ref{eq:G_full}) and (\ref{eq:G_asymm1}), and then
making use of Eqs.~(\ref{eq:split_op}) and (\ref{eq:G_trans}): 
\begin{eqnarray}
\tilde G (\vec{R}, \vec{\Omega} \rightarrow \vec{R'}, \vec{\Omega'}, \Delta\tau) 
	& 	\simeq   &
	\frac{1}{\left(  4\pi D \Delta\tau\right)^{3N/2} } \nonumber\\
& &
    e^{-(\vec{R'} - \vec{R}-D\Delta\tau \vec{F}(\vec{R}))^2/4D\Delta\tau } \,
    e^{-\Delta\tau[(V(\vec{R'},\vec{\Omega'}) + V(\vec{R},\vec{\Omega}) )/2 -E_R]} 
		\nonumber \\
& \prod_{p\alpha} &\langle \phi'_{p\alpha} | e^{-[\tilde{d}_{p\alpha} \hat{L}^2_{p\alpha} 
+ i \hbar \tilde{d}_{p\alpha} \hat{L}_{p\alpha} f_{p\alpha}]\Delta\tau} | \phi_{p\alpha} \rangle 
\label{eq:Gtilde_rot}
\end{eqnarray}
Inserting a complete set of one-dimensional angular momentum eigenstates,
Eq.~(\ref{eq:angmom_eigenf}), 
in each of the terms inside the product on the right hand side as before (Section~\ref{unbiased})
leads to, e.g., for 
$\alpha = x$,
\begin{equation}
\langle \phi'_{px} | e^{-[\tilde{d}_{px} \hat{L}^2_{ix} 
+ i\hbar \tilde{d}_{px} \hat{L}_{px} f_{px}]\Delta\tau} | \phi_{px} \rangle =
\frac{1}{2\pi} \sum_m e^{-d_{px} m^2 \Delta\tau 
+ i m (\phi_{px} - \phi'_{px} - d_{px} \Delta\tau f_{px}) },
\label{eq:onedim_bsd}
\end{equation}
with $d_{p\alpha} = \hbar^2 \tilde{d}_{p\alpha}$ as before.
This can be shown to reduce to a Gaussian form in the small 
$\Delta\tau$ limit, via the same procedure as employed in Section~\ref{unbiased}.  
Thus, replacing $\phi$ in Eq.~(\ref{eq:gauss_expand}) now by the angular displacement 
including the contribution from the angular quantum force, e.g., for $\alpha = x$,
one has $(\phi_{px} - \phi'_{px} -d_{px} \Delta\tau f_{px} )$, leads to the 
desired Gaussian form of the sum in Eq.~(\ref{eq:onedim_bsd}), i.e.,
\begin{eqnarray}
\frac{1}{2\pi} \sum_m e^{- d_{px} m^2 \Delta\tau 
+ i m(\phi_{px} - \phi'_{px} -d_{px} \Delta\tau f_{ix}) } 
& \stackrel{\Delta\tau \rightarrow 0}  {\simeq} & \nonumber \\
& & \hspace{-2cm}
 \frac{1}{\sqrt{4\pi d_{px} \Delta\tau}} 
e^{-(\phi_{px} - \phi'_{px} - d_{px} \Delta\tau f_{px})^2 / 4d_{px} \Delta\tau}.
\label{eq:equate_gauss_bsd}
\end{eqnarray} 
We thereby arrive at the full, short-time factorized, limiting form for the translational and
rotational importance sampled Green's function for a set of interacting rigid bodies:
\begin{eqnarray}
 \tilde G (\vec{R}, \vec{\Omega} \rightarrow \vec{R'} \vec{\Omega'}, \Delta\tau) 
	 	&\simeq  & 
\frac{1}{\left(4\pi D \Delta\tau\right)^{3N/2}} 
    e^{-(\vec{R'} - \vec{R}-D\Delta\tau \vec{F}(\vec{R}))^2/4D\Delta\tau }  \nonumber\\
& &   e^{-\Delta\tau[(V(\vec{R'},\vec{\Omega'}) + V(\vec{R},\vec{\Omega}) )/2 -E_R]}
\times\nonumber \\
\label{eq:Gtilde_full}
& 	\prod_{p\alpha}& \frac{1}{\left( 4\pi \Delta\tau d_{p\alpha} \right)^{1/2} }
   e^{-[(\phi_{p\alpha} -\phi'_{p\alpha} - d_{p\alpha} \Delta\tau f_{p\alpha})^2
/4d_{p\alpha}\Delta\tau]}
\end{eqnarray}
This may be generalized to multiple values of $D_p$, by replacing the first factor with
\begin{equation}
	\left( {1 \over 4\pi  \Delta\tau}\right)^{3N/2} 
	\prod_p \left( {1 \over D_p} \right) ^{3/2} e^{-(R'_p - R_p-D_p\Delta\tau F_p(R_p))^2/4D_p \Delta\tau }. 
\label{eq:trans_gen}
\end{equation}
Details of the implementation of Eq.~(\ref{eq:Gtilde_full}) 
are presented in the next Section.
 
We note that, for the specific case of diatomic molecules, Lewerenz has proposed
another method to implement rigid rotor motion, namely the method
of adiabatic constraints~\cite{lewerenz96}.  This method is
attractively simple and easy to implement for ground state studies of
linear molecules~\cite{lewerenz96,blume96,gianturco00}, although it 
becomes considerably more involved when applied to more general rigid bodies~\cite{suhm91}. 
Importance sampling of rotational degrees of freedom would be problematic within such an
approach however, since it is not clear how to combine this with the adiabatic constraints. 
In particular, it is not obvious at what stage the quantum forces should be introduced, 
i.e., before or after imposition of the adiabatic constraint.  
Recently, Sarsa {\it et al} have shown that classical constraint dynamics can be used to generally impose bond length and bond angle constraints, in both biased and unbiased DMC\cite{sarsa00}.

\section{Mixed frame and fixed frame implementations}
\label{implement}

In order to demonstrate Monte Carlo solution of Eq.~(\ref{eq:green}) with
Eqs. (\ref{eq:G_full}) and (\ref{eq:Gtilde_full}), 
we consider a system composed of 
$m$ rigid rotors and $n$ atoms ($n+m=N$). 
The Monte Carlo propagation over
a step $\Delta \tau$ requires then $3(m+n)$ translational diffusion moves, 
 $3m$ rotational diffusion moves, and the local exponential growth/decay 
operations.  
The translational Green's function, Eq.~(\ref{eq:G_trans}), consists of a product of a diffusive term deriving from $\hat T^t$, the translational kinetic energy, and a branching term deriving from the potential energy, $V$. 
Two kinds of moves are thereby made on an initially established
ensemble of multi-dimensional configuration points or ``walkers'', 
Eq.~(\ref{eq:initial}).  
The first kind of move is a diffusion from
$\vec{R}_j$ to new positions $\vec{R}'_j$, implemented by sampling the 
vector $\vec{R}_{j} - \vec{R'_j}$ from a $3N$-dimensional 
Gaussian distribution, of width $\sqrt{2D_p\Delta\tau}$ in each of the $3N$
cartesian coordinates.  The second move is a ``branching'' which 
modifies the
weight of each walker in the ensemble by the exponential factor
$e^{-\Delta\tau[(V(\vec{R'}) + V(\vec{R}) )/2 -E_R]}$.  This can be
implemented by i) replicating/destroying configurations, or by ii)
maintaining and updating a continuous weighting factor associated with
each configuration, or by iii) a combination of weights and
replication\cite{hammond94,blume96}.
The role of the reference energy $E_R$ now becomes clear, 
i.e., it provides a means to reduce the fluctuations deriving from the 
branching term.

Propagation with $\tilde{G}^t$, Eq.(\ref{eq:Gtilde_trans}), is performed by a simple modification of the
unbiased propagation described above, namely that the
diffusive moves are modified to include the contribution from the quantum
force $\vec{F}(\vec{R})$.  The coordinates of each particle are therefore
updated according to, e.g., for $R_{px}$,
\begin{equation}
    R'_{px} =  R_{px} +\Delta_{px} + 2 D_p \Delta\tau \left( { \partial \psi_T(\vec{R}) 
\over \partial x_p} \right) / \psi_T(\vec{R})  
\label{eq:trans_move}
\end{equation}
where $\Delta_{px}$ is a random number sampled from a Gaussian of
width $\sqrt{2D_p\Delta\tau}$.
The associated modification of the branching term consists in  using the 
local energy instead of the potential, as shown in the 
exponential factor\linebreak[4] $e^{-\Delta\tau[(E_L(\vec{R'}) + E_L(\vec{R}) )/2 -E_R]}$\nolinebreak[4] for the update of the weights and/or the replication of walkers.

The implementation for the rotational Green's function, unbiased $G^r$  and 
biased $\tilde G^r$, is similar.
In the unbiased formalism, diffusive rotational moves consist in sampling the three rotation angles
for each rotor from a $3m$-dimensional Gaussian distribution of width $\sqrt{2d_{p\alpha}\Delta\tau}$.
In the biased version, these rotation angles also include the quantum force :
\begin{equation}
    \phi'_{p\alpha} = \Delta_{p\alpha} + 2 d_{p\alpha} \Delta\tau \left( { \partial \psi_T(\vec{R},\vec{\Omega})
\over \partial \phi_{p\alpha}} \right) /\psi_T(\vec{R},\vec{\Omega}).
\label{eq:rot_move}
\end{equation}

For the full Green's function, both translational and rotational diffusion
moves are made. We now give a more detailed explanation of the various
possible implementations of this in different frames of reference.

\subsection{Mixed Frame}
\label{mixed}
In the general case, {\it i.e.} for more than one rotor, it is necessary to use different frames for the rotation and the translation moves.
The rotation must be performed in the principal axes frame, PAF of each rotor, the translations are done in the laboratory frame, SF.
The evaluation of the potential and of the trial function have to be done in consistent frames.
At each time step in the random walk, three rotations around the principal axes of the molecule are made.
During the time step $\tau-\Delta\tau  \rightarrow \tau $, the PAF  (PAF$_{\tau-\Delta\tau}$) is thus rotated into a
new PAF: (PAF$_{\tau}$). 
The rotation matrix describing the
attempted move is :
\begin{equation}
\label{Rmatrix_prod}
{\hat R}^{(\tau)} = {\hat R}_x(\phi_x^\tau) \cdot {\hat R}_y(\phi_y^\tau)
\cdot {\hat R}_z(\phi_z^\tau)
\end{equation}
where ${\hat R}_\alpha(\phi) $ represents a rotation of $\phi$ around the $\alpha$-axis, and $\phi_\alpha^\tau$ denotes the angle moved around this axis in time $\tau$. 
The corresponding relation between the coordinates in the new PAF and in the old 
can be derived using the following identity:
\begin{equation}
\label{Rot_move}
{\hat R}^{(\tau)}={\hat R}_{z^{''}}(\phi_z^\tau) \cdot {\hat R}_{y^{'}}(\phi_y^\tau) \cdot {\hat R}_x(\phi_x^\tau).
\label{eq:R_indep}
\end{equation}
Here $y^{'}$($z^{''}$) is the new orientation of the $y$ ($z$) axis after the first (the first and second) rotation(s).
This identity is easily verified
by transforming each of the rotations appropriately~\cite{zare}, e.g.
\begin{equation}
{\hat R}_{y^{'}}(\phi_y^\tau) = {\hat R}_x(\phi_x^\tau) \cdot
{\hat R}_{y}(\phi_y^\tau) \cdot {\hat R}^T_x(\phi_x^\tau).
\end{equation}

The matrix representation of Eq.(\ref{Rot_move}) is given by:
\begin{eqnarray}
\left(
\begin{array}{c}
x\\
y\\
z\\
\end{array}
 \right)_{PAF_\tau} & = & \left(
\begin{array}{ccc}
\cos \phi_z &   \sin \phi_z  &0 \\
- \sin \phi_z & \cos \phi_z  &0 \\
 0 & 0 & 1
\end{array}
\right)  \left(
\begin{array}{ccc}
\cos \phi_y &0&  - \sin \phi_y   \\
 0 & 1 & 0 \\
\sin \phi_y &0& \cos \phi_y
\end{array}
\right) \nonumber\\
& &
  \left(
\begin{array}{ccc}
1 & 0 & 0 \\
0 &  \cos \phi_x & \sin \phi_x   \\
0 & - \sin \phi_x & \cos \phi_x
\end{array}
\right) \left(
\begin{array}{c}
x\\
y\\
z\\
\end{array}
 \right)_{PAF_{\tau-\Delta\tau}} \label{Rmat} \\
\left(
\begin{array}{c}
x\\
y\\
z\\
\end{array}
 \right)_{PAF_\tau} & = &
{\mathcal R}^{(\tau)} (\phi_z,\phi_y,\phi_x)
\left(
\begin{array}{c}
x\\
y\\
z\\
\end{array}
 \right)_{PAF_{\tau-\Delta\tau}}.
\label{Rmatrix}
\end{eqnarray}

If the initial orientation of the PAF is given by 3 Euler angles $(\varphi_0, \vartheta_0, \chi_0)$ with $\varphi_0 \in [0,\pi)$, $\vartheta_0 \in [0,\pi]$ and $\chi_0\in [0,\pi)$,  Eq.  (\ref{Rmatrix}) can be used to compute the relation between the coordinates in the PAF and in the SF. Then
\begin{eqnarray}
\left(
\begin{array}{c}
x\\
y\\
z
\end{array}
 \right)_{PAF_\tau} &= &
\prod_{i=1}^n {\mathcal R}^{n \Delta\tau} \ldots {\mathcal R}^{2 \Delta\tau}
 {\mathcal R}^{\Delta\tau} {\mathcal R}^0
\left(
\begin{array}{c}
x\\
y\\
z
\end{array}
 \right)_{SF} =
{\mathcal P}(\tau)
\left(
\begin{array}{c}
x\\
y\\
z
\end{array}
 \right)_{SF},
\end{eqnarray}
with
\begin{eqnarray}
{\mathcal R}^0 & = & \left(\begin{array}{c}
\cos \varphi_0\; \cos \vartheta_0\; \cos \chi_0 - \sin \varphi_0\; \sin \chi_0\\
-\cos \varphi_0\; \cos \vartheta_0\; \sin \chi_0 - \sin \varphi_0\; \cos \chi_0\\
\cos \varphi_0\;\sin \vartheta_0
\end{array} \right.  \\
& & \hspace{1.5cm} 
\left.\begin{array}{cc}
 \sin \varphi_0\; \cos \vartheta_0\; \cos \chi_0 + \cos \varphi_0\; \sin \chi_0
& - \sin \vartheta_0\; \cos \chi_0\\
 -\sin \varphi_0\; \cos \vartheta_0\; \sin \chi_0 + \cos \varphi_0\; \cos \chi_0
&  \sin \vartheta_0\; \sin \chi_0\\
 \sin \varphi_0\; \sin \vartheta_0 
&  \cos \vartheta_0
\end{array}\right). \nonumber
\end{eqnarray}

In an unbiased ``pure'' DMC calculation,
$\phi_x, \phi_y$ and $\phi_z$ are sampled from a
Gaussian distribution which is centered at zero and has
width $\sqrt{2 d_\alpha \Delta\tau}$. In an
importance sampled calculation, we have to add the quantum force,
according to Eq.~(\ref{eq:rot_move}).
Rotation around $x$ modifies the orientation of the $y$ principal axis of the rotor.
The second rotation is thus made around the new orientation of this axis,
followed by the third rotation around the subsequent new orientation of the
$z$ principal axis.
Due to the small values of these angles, we neglect the non-commutation 
of these 3 rotations.
	
In the Mixed Frame formulation, the SF cartesian coordinates for the centers
of mass of all particles are used at every time step in the scheme.
The use of the $\mathcal P$ matrix allows us to compute the coordinates of the 
relative vectors between the atoms and the rotors, in the PAF of each rotor. 
These relative vectors are needed for evaluation of the potential term and of
the trial wave function.

As noted above, the Mixed Frame formulation is needed whenever the system 
contains more than one rotor. 
It is also required 
whenever the Euler angles defining the orientation of the PAF in the SF are 
needed.  These angles are defined by the $\mathcal P$ matrix.
Knowledge of these  angles is 
necessary for  calculation of excited states within a fixed node
approximation whenever the nodes are imposed in a SF frame, because the 
trial function is then an explicit function of these angles.
In the particular case of a trial function that depends only on the 
second Euler angle , e.g. 
$\Psi_T = |1,0,0\rangle  \propto \cos \vartheta = {\mathcal P}_{33}$, 
computation and storage of only the last column of $\mathcal P$ is sufficient.
Section \ref{subsect_HCN} provides an illustrative example of calculation of
internal bending excitations for He-HCN using this Mixed Frame formulation.

\subsection{Fixed Molecular Frame}
\label{mframe}
When the system under consideration contains only one rotor, the Monte Carlo propagation can be performed in the PAF
 of this rotor. 
Like in the previous section, at each time step in the random walk, three rotations around the principal axes of the molecule are made.
The corresponding relations between the coordinates are given by equation (\ref{Rmatrix}).

In order to perform the translational moves, one continuously follows the molecular orientation.
Then at each  time step ($\tau$), the atoms and the center of mass of the rotor are defined by their coordinates in the molecular PAF$_\tau$.
The matrix ${\mathcal R}^{(\tau)}$ is used to compute the  coordinates in the new PAF using their coordinates in the former one (PAF$_{\tau-\Delta\tau}$).
The translational moves are then made according to Eq.~(\ref{eq:trans_move}),
using the coordinates in the old PAF.  
Vectors describing the position of the atoms in the two PAF's are used to 
compute the atom-rotor interaction potential
and the trial wave function, before and after the step.

Calculations for SF$_6$-He system using the implementation described in this section leads to the same results as 
those obtained with the Fixed Laboratory Frame. One disadvantage of this 
Fixed Molecular Frame scheme is that it does not allow for easy computation 
of the Euler angles defining the orientation of the PAF with 
respect to the SF frame.

\subsection{Fixed Laboratory Frame}
\label{lframe}
In the case of a spherical top molecule, Eq. (\ref{eq:L}) is applicable in 
the space-fixed laboratory reference frame (SF) as well as in the PAF, 
because 
{\it any} choice of orthogonal frame diagonalizes the inertia tensor.
The rotational moves can therefore be made around the axes of the 
SF frame ($X,Y,Z$).
Although the rotation moves can be made without knowing the orientation of
the rotor at a given time,
it is necessary to keep track of the molecular orientation in the SF frame
because
computation of the interaction with the other rotors and atoms (Eq.~(\ref{eq:V_pair})) requires
knowledge of the orientation of
the preassigned reference directions of the spherical top (e.g. the
S---F bonds in the case of SF$_6$ He$_n$ clusters).

Rotational moves, performed according to Eq.~(\ref{eq:R_indep}), are combined with 
translational moves of the $n$ atoms and of the $m$ center of masses of the 
rotors, made in the cartesian coordinates of this Laboratory 
Frame (SF), within the conventional approach (Section~\ref{implement} above).
This fixed Laboratory Frame scheme has
recently been implemented in a study of rotational excitations of
SF$_6$ in helium clusters~\cite{lee99}, and is also used in the
SF$_6$ applications presented here.

\section{Applications}
\label{applications}

\subsection{ Angular Sampling in Three Dimensions: SF$_6$-He}
\label{1d3d}

The SF$_6$ He$_n$ system with $n=1$ was used to test the rotational 
sampling algorithm. In particular, since for a spherical top we know 
the exact three-dimensional rotational Green's function (Eq.~(\ref{eq:G_sph}) 
and Appendix A), we can check the accuracy of the
short-time decomposition of this into three one-dimensional rotational
factors, Eq.~(\ref{eq:G_rot}).  
The calculations are carried out in the
Fixed Laboratory Frame as described in section~\ref{lframe}, and the SF$_6$-He
interaction is modeled using the anisotropic potential of 
Pack et al\cite{pack84}.
We implemented both sampling schemes for this system, i.e.,
one rotation about a randomly chosen axis, corresponding to
Eq.~(\ref{eq:G_sph}), and three rotations about the 
space fixed $X, Y$, and $Z$ axes, corresponding to Eq.~(\ref{eq:G_rot}).  
Figure \ref{Figure_euler} (dotted lines) shows the evolution of the 
distribution of Euler angles (orientation) 
obtained from the direct sampling of the three-dimensional rotational Green's 
function.  It also shows (solid lines) the evolution of the Euler 
angle distributions obtained with the factorization into three 
independent rotations.  Both of these methods yield similar distributions
that exhibit the expected Gaussian form (see Appendix A).
The computed ground state energies are summarized in 
Table~\ref{table_n1_energies}.
\begin{table}[t]
\caption[]{Unbiased DMC energies in K for the SF$_6$-He system.
The numbers in parentheses give one standard deviation statistical errors
in the last digits.
$^*$The FBR/DVR value is converged, {\it i.e.}, it is stable when the
unsymmetrized basis set size is increased from 19890 to 33930.}
\begin{tabular}{ccc} \hline
\hspace{1cm}FBR/DVR \hspace{1cm} & 
\hspace{1cm}DMC \hspace{1cm} & 
\hspace{1cm}DMC (no rotations)\hspace{1cm}  \\ \hline
-37.14$^*$ & -37.4(2) & -38.4(3) \\\hline
\end{tabular}
\label{table_n1_energies}
\end{table}         
\begin{figure}[b]
\vspace{18cm}
\caption[]{Euler angle distributions for SF$_6$ calculated using the three
rotations about fixed cartesian axes scheme (solid lines),
and using  the single rotation about a randomly chosen axis scheme
(dotted lines). The initial configurations all started with the same
arbitrary initial molecular orientation.  The distributions shown here
are plotted at three increasing time slices to demonstrate the gaussian spreading.}
\label{Figure_euler}
\end{figure}       
The value obtained with rotation incorporated is in good agreement with 
result obtained using basis set expansion methods.
These reference calculations employed 
the Finite Basis Representation - Discrete Variable Representation
(FBR-DVR) collocation scheme of Leforestier using a Wigner 
basis set~\cite{leforestier94}.  We also show the energy obtained when the molecular rotation is omitted.  This is ~$\sim$ 1~K lower, and is in good
agreement with previous DMC studies made without rotation~\cite{barnett93}. 
Since the energetic effect of rotation is very small for $n=1$, i.e.,
the rotational energy increment in the SF$_6$-He ground 
state is only $\sim~$1 K,
we prefer to investigate the accuracy of the  two different short time Green's
function approximations using an artificial SF$_6$ having rotational
constant $B_0 = 0.91$ cm$^{-1}$, i.e., 10 times the true gas phase value.
This increases the variation of the energy due to rotation and amplifies 
the effect of any approximations. Rotating by performing a single rotation
about a randomly chosen axis yields a ground state energy 
(E$_{\rm 1rot}$ = -33.9(6) K) 
that is consistent with both the value obtained by implementing the 
three rotations about fixed axes scheme 
(E$_{\rm 3rot}$ = -33.8(2) K) 
and a FBR-DVR reference calculation 
(E$_{FBR}$ = -33.6 K).
In contrast, incorrect sampling (omitting the $a = 3$ factor discussed in
Appendix A) yields erroneous energies which are too low 
(E = -35.8(7)).  
This omission is operationally equivalent to reducing the rotational constant by a factor of 3.

\subsection{Importance sampling to overcome artificial dissociation in 
medium-sized weakly bound clusters} 
\label{structures}

The necessity of importance sampling appears already in ground state studies
of moderately sized, weakly bound clusters, e.g., SF$_6$ He$_n$ and 
HCN He$_n$ with $n>12$.
Both of these systems require importance
sampling to avoid the unphysical dissociation of helium atoms which 
can result from the occasional diffusion of He out to large distances from 
the cluster.  In unbiased DMC, once an atom diffuses very far from the 
weakly bound cluster it has only a very small probability to return
diffusively to the region of stronger binding, while the energetic penalty
for motion further away from the molecule becomes negligible.
Moreover the configuration space for a motion toward the molecule 
is much smaller than the one for a motion away.
  This is
therefore a metastable situation from which it is very difficult to return
to configurations of strong binding. 
The result is that helium atoms become "lost" outside the boundaries 
of the simulation box, leading to false dissociation as well as to
artificially high energies. 

One of the attractive features of unbiased DMC is that it 
provides the full-dimensional wave function.  This may be fit to obtain trial 
functions for importance sampling, as has been done by a number of 
authors~\cite{mccoy99,barnett92,broude95}. 
For the molecule-helium clusters with $n=1$, the helium wave function may be
squared to obtain the density.
However, for $n>1$, three-dimensional representation of the $n$-helium wave 
function is  not possible. In order to nevertheless provide a visual 
representation of the wave function, we shall
extract an effective one-particle wave function by projection, binning 
independently the position of the $n$ heliums in the molecular frame.

For the spherical top SF$_6$ molecule, these calculations are performed in 
the space-fixed, laboratory frame (SF), described in the previous section.
For  the linear HCN molecule, calculations are performed in the Mixed
Frame scheme of Section~\ref{mixed}.  Rotation around the $z$-axis of the PAF 
is not allowed ($B_z=0$), and 
consequently, the rotation formalism presented there is reduced to two 
angles.

Simple trial wave functions can be used to avoid artificial dissociation.  
For one rigid molecule, a form of trial wave function 
possessing the correct permutation symmetry is given by 
the product of pair correlation terms\cite{whaley98}:
\begin{equation}
\Psi_T(\vec{R},\vec{\Omega})  = 
\prod_{p=1}^{n}\Psi_{He-X}(R_p,R_X,\Omega_X) 
\prod_{p\neq q}^n\Psi_{He-He}(R_p,R_q) 
. 
\end{equation}
In the examples presented here for $N \leq 20$, the helium-helium correlation 
is relatively unimportant compared to the helium-molecule term, so for 
simplicity we have used $\Psi_{He-He}(R_p,R_q)=1$.
For SF$_6$He$_{1}$ and SF$_6$He$_{20}$, the helium atoms lie within
the first solvation shell. 
As long as we are interested primarily in the binding energies here, and not
concerned with detailed structural information in the angular degrees of
freedom, we can adequately describe the He-SF$_6$ correlations using 
 simple radial wavefunctions, of the form 
\begin{equation}
\Psi_{He-X}(R_p,R_X,\Omega_X) = \exp\left[- c r_{pX}^{-5} - a r_{pX}\right], 
\label{sf6hewave}
\end{equation}
where $r_{pX}=|R_p-R_X|$, and $X$=SF$_6$. 
For this demonstration with $n=20$, we used $a=0.80$ and $c = 42000$ in atomic units. These
parameters were obtained from fitting results of unbiased simulations for similar $n$ values. Care
should be taken not to choose a trial wavefunction that is too restrictive
(i.e. negligible amplitude in regions that may actually be important). Such
choices of wavefunction can give rise to instabilities and to erroneous
energies. 
If the radial trial function is too narrow, falling off too
sharply in the repulsive regime close to the impurity, 
the convergence of the DMC scheme becomes more difficult
and the energy can become stuck 
below the true value.  We found this to be the case when a 
trial function fit to unbiased wave functions obtained for small $n$ 
($n=1$ with $a=1.5$, $c=6000$ a.u.) was
employed for importance sampling of a cluster with significantly larger $n$
($n=20$).  This
emphasizes the importance of choosing a trial function which does not impose
excessive restrictions on the sampling. 



For HCN, we used the radial part obtained from the high quality anisotropic
He-HCN trial function derived for use in the fixed node calculations of 
Section V.C.  This is of the form
\begin{equation}
\Psi_T(R_p,R_X,\Omega_X)  =  \exp\left[c_{10}r_{pX}^{c_{20}}
+ c_{30}\ln(r_{pX}) - \exp ( c_{40}-c_{50}r_{pX}) \right],
\end{equation}
where $r_{pX}=|R_p-R_X|$ with $X$=HCN, and 
parameters presented in Table \ref{Table_HCN_para}.
\begin{table}[h]
\caption{Parameters (in au) used for the definition of the HCN-He trial wavefunction.}
\begin{tabular}{crrrrrr} \hline
       & \hspace{1cm}$a_i$ & \hspace{1cm}$c_{1i}$ & \hspace{1cm}$c_{2i}$& \hspace{1cm}$c_{3i}$& \hspace{1cm}$c_{4i}$& \hspace{1cm}$c_{5i}$ \\ \hline
$\ell=0, i=1$& 1.00 &20.07 &-0.051  & -9.18 &5.38  & 0.61   \\
$\ell=1, i=1$& 1.00 &25.96 &-0.055  & -11.39&5.66  & 0.58   \\
$\ell=2, i=1$& 2.90 &20.17 &-0.031  & -10.14&6.52  & 0.74  \\
$\ell=2, i=2$& -0.43&16.38 &0.035  & -9.33&6.30  & 0.85  \\
$\ell=3, i=1$& 4.61 &26.77 &-0.029  &-13.40&6.68  & 0.72  \\
$\ell=3, i=2$&-6.40 &37.37 &-0.059  &-17.00&5.44  & 0.53 \\ \hline
\end{tabular}
\label{Table_HCN_para}
\end{table}     

The effective one-particle helium wave functions obtained by unbiased RBDMC 
are shown in Figure \ref{Figure_SF6_HE20}a for SF$_6$He$_{20}$,
and in Figure \ref{Figure_HCN_HE20}a for HCN He$_{20}$. 
The "wings" at the outer edge of the binning domains are the signature of 
dissociating helium atoms.  They appear here because we put all walkers 
having a position larger than the domain definition into the last bin
at the edge of the domain. The average number of artificially dissociated 
atoms defined as those lying at a distance greater than 20 a.u. for both systems,
is $n_{diss}\sim 5$ for HCN, and $n_{diss}\sim 1$ for SF$_6$.
\begin{figure}[t]
\caption[]{(a) Radial wavefunction profile for SF$_6$ He$_{20}$ obtained using unbiased DMC and projecting the helium wave function into the molecular frame.
This is arbitrarily normalized as if it were a density.
(b) The mixed helium density $<\Psi_T|\rho |\Psi>$ for SF$_6$ He$_{20}$
obtained using a radial trial function (see text).
}
\vspace{15cm}
\label{Figure_SF6_HE20}
\end{figure}             
\begin{figure}[t]
\caption[]{(a) One-particle effective ground state wave function for HCN $^4$He$_{20}$, obtained using unbiased DMC and projecting the helium wave function
into the molecular frame.
The HCN molecule lies on the $z$ axis, with its center of mass at the origin and the H
atom on the $z>0$ side. The second axis ($r$) is the distance to that axis.
The non-physical peaks at the edges of the box come from the binning procedure and show artificial dissociation of some helium atoms - see text.
(b) Representation of the mixed density $<\Psi_T|\rho |\Psi>$  obtained using a radial
trial function (see text).}
\vspace{17cm}
\label{Figure_HCN_HE20}
\end{figure}        

This dissociation phenomenon also has a pronounced effect on the energy.
For clusters containing smaller numbers of helium atoms, unbiased RBDMC and
importance sampled DMC lead to the same energies, as they should 
if both calculations are converged. For $n=20$, convergence of the 
importance sampled DMC is straightforward, yielding a ground state energy
of -237(4) K
for HCN and -607 (11) K for SF$_6$. 
However with unbiased DMC, in sharp
contrast to this situation,  full convergence of the 
energy is much more difficult - even impossible - for this number of helium 
atoms,
because of the artificial dissociation of helium walkers,  
for both HCN and SF$_6$.
We can only obtain an upper estimate of the ground state energy as a result.
For SF$_6$ this estimate is $\sim$ -564 K, 
and for  HCN it is 
$\sim$ -196 K.
%
In both cases the unbiased DMC value is considerably higher than the
fully converged importance sampled value, reflecting the smaller number of 
truly bound helium atoms in the unbiased calculations.  It is significant
that this phenomenon is seen not only for the relatively weakly bound HCN
molecule in $^4$He$_n$, but also for the much more strongly bound SF$_6$ 
molecule.
Our analysis implies that this is a general phenomenon which will occur for any
species beyond a minimum size which appears to be somewhat less than one solvation 
shell. 

Figures \ref{Figure_SF6_HE20}b and \ref{Figure_HCN_HE20}b show the 
mixed densities $<\Psi_T|\rho |\Psi>$ derived from the simple radial trial 
functions with importance sampled RBDMC, for SF$_6$ and  for HCN, respectively. 
It is evident by comparison with 
Figures \ref{Figure_SF6_HE20}a and \ref{Figure_HCN_HE20}a that the use of a simple radial function which provides
binding of
the helium atoms to the embedded molecule very effectively removes the 
artificial dissociation 
problem, and allows us to fully converge a ground state energy calculation.
Although we do not evaluate structural quantities here other than this mixed
density, we note that the use of importance sampling does allow a relatively
straightforward access to
computation of structural quantities using second-order 
estimators~\cite{whaley98}.

Table \ref{summary} summarizes the energies and the number of artificially 
dissociated helium atoms for both clusters.
\begin{table}[h]
\caption{Comparison between unbiased and biased RBDMC for SF$_6$ and HCN with 20 helium atoms. No error estimates are shown for the unbiased results, because the convergence
with this approach is poorly defined (see text).}
\begin{tabular}{cccc} \hline
               & Unbiased energy in K & $n_{diss}$ & Biased energy in K \\ \hline
SF$_6$ He$_{20}$& -564 & $\sim$1 & -607 (11)             \\
HCN He$_{20}$ & -196   & $\sim$5 & -237 (4) \\\hline
\end{tabular}
\label{summary}
\end{table}         

\subsection{Excited States} \label{subsect_HCN}
\label{nodes}

Our third application addresses the computation of excited states for the 
He-HCN complex.
Experiments in doped helium clusters show that the rotational 
spectrum of small molecules inside clusters of $^4$He possesses 
the same symmetry as the corresponding spectra in the gas phase, but that 
the effective 
rotational constants are reduced~\cite{toennies98,kwon00}.
For HCN this reduction is about 20\%,\cite{nauta99,conjusteau00} 
considerably less than the ~$\sim$ 80\% reduction seen for the more strongly 
bound species such as SF$_6$~\cite{hartmann95}.
Table \ref{Table_d} shows that HCN is much more weakly bound to helium than SF$_6$, but
that in both cases the ground state energy of the $n=1$ complex lies
above the saddle point of the interaction potential.
Analysis of such rotationally excited states within a fixed node 
approximation requires some estimate of the relevant nodal structure.
Insight into this can be obtained from level assignments made according
to simple models.
The earlier study of SF$_6$ in $^4$He$_N$ from our group calculated 
rotational 
levels corresponding to $J \sim j$, where $J$ is the total angular momentum, 
and $j$ the molecular angular momentum~\cite{lee99}.
For He-HCN, we evaluate here instead the excited state 
which has been assigned by Atkins and Hutson~\cite{atkins96}, and by
Drucker et al~\cite{drucker95}, to the level 
$|j=1, \ell=1, \bf{J=0} >$. 
In this assignment scheme, $J$ is the total 
rotational quantum number, $j$ 
the quantum number of HCN, which is to a good approximation conserved in the 
weakly bound complex, 
and $\ell$ an ``orbital'' quantum number associated with the rotation of the 
helium around the HCN.  
For this excited state the fixed nodal approximation can be imposed in the
molecular frame, as we describe below.  
Nevertheless,
we employ here the Mixed Frame implementation of the rotational importance
sampling algorithm, rather than the Fixed Molecular Frame
version.  For a linear molecule, while we must always perform the 
rotations in the molecular frame, nodal structures may be 
imposed in either the MF or the SF frame, leading to the
natural choice of the Mixed Frame implementation as the most general in this
case.  

In order to obtain a trial wave function for the fixed node
calculations, we first performed pure DMC runs
which allow us to compute the ground state eigenfunction 
(see Figure \ref{Figure_HCNHe}).
\begin{figure}[h]
\caption[]{Ground state helium wave function obtained by pure DMC propagation projected into the molecular frame for  He-HCN.
The orientation of the HCN molecule is the same as in figure \ref{Figure_HCN_HE20}.}
\vspace{8cm}
\label{Figure_HCNHe}
\end{figure}                 
This was fit to an analytic form 
$\Psi_T(r,\theta)$, 
where $r$ and $\theta$ are the usual Jacobi coordinates of the He with respect 
to the center of mass and orientation of HCN, to obtain a high quality trial function.  We 
use the following Legendre polynomial expansion:
\begin{eqnarray}
\Psi_T(r,\theta) & = & \sum_{\ell=0}^3 \Psi_\ell(r) P_\ell(\cos \theta) \nonumber \\
\Psi_\ell (r) & = & \sum_{i=1}^2 a_i\exp\left[c_{1i}r^{c_{2i}}
+ c_{3i}\ln(r) - \exp ( c_{4i}-c_{5i}r) \right]\nonumber
\end{eqnarray}
with $\left\{a_i , c_{ki}, k=1,\ldots,5 ; i=1,2\right\} $ the parameters of the fit presented in Table \ref{Table_HCN_para}.
Limiting the sums to $\ell \leq 3$ and $i \leq 2$ does not 
noticeably
affect the precision of the fit.

Such a high quality trial function is necessary for the fixed node 
calculations of the He-HCN complex described here.  
In order to construct a trial excited state function for a fixed node 
calculation, 
we combined this high quality ground state trial function with a simple
function that imposes the nodal constraint. This is chosen here by
comparison with the exact excited state function, which was obtained by
repeating the collocation calculations of Aktins and Hutson\cite{atkins96} 
to compute and analyze the wavefunction for this level. The energy
of this level is 
$-8.5$ K, 
and 
a contour plot of the corresponding
wave function is shown on figure \ref{Figure_HCN_He_DVR}. 
\begin{figure}[h]
\caption{Excited state $|1 1 \bf{0} \rangle$ wavefunction obtained by DVR calculation represented as a function of the Jacobi coordinates $r$ in \AA$\;$  and  $\theta$ in degrees. The contour lines are evenly spaced ranging from -0.5 to 0.4}
\vspace{8cm}
\label{Figure_HCN_He_DVR}
\end{figure}              
The structure of this function implies a strong internal bending character
to this excitation.
The excited state function possesses a single nodal surface, which is seen 
to be approximately $r$-independent.
This motivated us to employ 
a fixed nodal surface defined by $\cos (\theta + \chi)$, 
where $\chi$ is a 
parameter and $\theta$ is the internal Jacobi angle of the cluster
{\it i.e.} the trial function used is thus 
$\tilde{\Psi}_T(r,\theta)=\Psi_T(r,\theta)\cos (\theta + \chi)$.
Since the calculation is carried out in the mixed frame formulation, this
trial function is implicitly dependent upon the orientation of the rigid
body in the space fixed frame, even though the nodal structure is imposed
in the molecular frame.
Comparison of the energy obtained for calculations restricted to each of the
two sides of the nodal surface allows us to then obtain the optimal value
of the parameter $\chi$.  This is arrived at when these two energies are
equal.  A similar procedure was used by McCoy and co-workers in an 
interactive sense to optimize nodal surfaces\cite{mccoy99}.  
For both the ground state and excited state computation, we use an ensemble of approximately 5000-6000 walkers.
We perform a block averaging in order to reduce the correlation between successive steps.
After equilibration of the ensemble, we perform 1 run of 800 blocks, each
of which consists of 150 time steps with $\Delta\tau =10$ au.
We reject all moves corresponding to a crossing of the nodal surface, keeping
the corresponding walker at its former position\cite{reynolds82}.

The optimal value of $\chi$ was found to be
$\chi=15.65$ degrees, which yielded  a common energy of 
-8.4(2) K 
on both sides of the resulting nodal surface. 
This fixed node value is in excellent agreement with the value 
-8.48797 K (-5.89897 cm$^{-1}$) obtained from the collocation method in 
Ref.~\cite{atkins96},
especially when the approximate nature of the nodal constraint is taken into
account. 

When studying clusters with larger numbers of
helium atoms using unbiased RBDMC, we noticed that the
$\theta$ dependence apparent in Figure \ref{Figure_HCNHe} tends to
smooth out~\cite{viel01}.
This suggests that for larger clusters, it might be adequate for some
purposes to reduce the trial function to the radial term only, e.g., for
studying ground state energies (Section V.B).

\section{Conclusions}

We have derived an algorithm for Rigid Body Monte Carlo with importance 
sampling for {\it all} degrees of freedom, including 
both translation and rotation, and
provided a rigorous derivation of the short time rotational Green's function
and its associated quantum forces.  
Three possible implementations of the algorithm were presented, which differ
according to the combination of frames used for rotational and translational
moves during the propagation.  The most general implementation scheme is
the mixed frame implementation in which translational moves are
made in the laboratory reference frame, while rotational moves of all rigid 
bodies are made in the principal axis frame of each monomer.  This 
allows full importance sampling calculations of molecular clusters to be
performed.

We have presented several applications of this importance sampled Rigid Body
Diffusion Monte Carlo. 
First, we demonstrated the 
correctness of the short-time factorization of the rotational
Green's function by comparison of the resulting product of one-dimensional
rotations with the well known three-dimensional formulation for a spherical
top.  
Second, we showed that importance sampling of translational degrees of
freedom is essential to avoid non-physical dissociation of weakly bound helium
atoms in doped quantum clusters of helium when more than a few helium atoms
are present.  
Third, we showed how excited states can be easily accessed with this
algorithm using the fixed node approximation and a trial wavefunction with
an implicit dependence on the orientation of the rigid body.  
The same algorithm may now
be applied to study excitations beyond the fixed node approximation, e.g.,
using the POITSE approach\cite{blume99}.

\section{Acknowledgements}
Financial and computational support from the National Science Foundation
through NSF grants CHE-9318737, CHE-9616615, and the NSF NPACI
program is gratefully acknowledged.  We thank Kevin Schmidt for pointing out
the need for the factor of three in the Gaussian sampling of angles for a 
spherical top (see Appendix A).  We thank Claude Leforestier for permission to
use his FBR-DVR Wigner Basis code\cite{leforestier94}.

\section{Appendix}
\subsection{Sampling of the Spherical Top Green's Function}
In order to provide a sampling of the operator
\begin{equation}
e^{-d\hat{L}^2\Delta \tau}
\label{app:op}
\end{equation}
where $d\equiv 1/2I$, according to the configuration representation of 
Eq.~(\ref{eq:G_sph}), one must choose both an arbitrary axis of rotation, and
an angle of rotation.  In order to implement this, it is necessary to first
specify the frame of reference of the operator $\hat{L}$, i.e., its axis of
quantization for $L_z$.  This is evident on expansion of the operator in
the angle space,
\begin{equation}
e^{-d\hat{L}^2\Delta \tau} = \int \frac{d \hat{\Omega}}{4\pi} e^{-ad(\vec{L}\cdot\hat{\Omega})^2\Delta \tau}
\label{app:expand}
\end{equation}
where $\hat{\Omega}$ is a unit vector in an arbitrary direction, and $a$ is 
a constant to be determined.
Expanding both sides of Eq.~(\ref{app:expand}) to order $(\Delta \tau)^2$, 
we find
\begin{equation}
1 - d L^2 \Delta \tau = 
	1 - \frac{a \Delta \tau}{4\pi} \int d\Omega (\vec{L}\cdot\hat{\Omega})^2.
\label{app:2nd_order}
\end{equation}
The integral can be evaluated once the direction of quantization of $\vec{L}$ is established.  Choosing this for simplicity as $\hat{z}$, so that 
$\vec{L}\cdot\hat{\Omega}=L\cos\theta$, and evaluating the integral over the 
spherical
angular coordinates $d\Omega \equiv d \phi d\theta$, leads to the equality
\begin{equation}
1 - d L^2 \Delta \tau = 1 - d\frac{a}{3} L^2 \Delta \tau,
\label{app:integral}
\end{equation}
from which we conclude that $a=3$.  Sampling of Eq.~(\ref{app:expand}) can 
then proceed by a) choosing a random vector $\hat{\Omega}$, and then
b) sampling the angle of rotation by choosing a value from
a gaussian of width $\sqrt{3\times (2d\Delta\tau)}$. This procedure provides
access to the full angular space in Eq.~(\ref{app:expand}).  When we 
are dealing with small rotations, this approach gives rise to similar
orientational distributions to the scheme in which 3 small rotations are
performed about the 3 cartesian axes (see Figures~\ref{Figure_euler}).
  This similarity will continue to hold as long
as the rotation $\zeta$ is small enough such that the single rotation 
about the arbitrary axis $\hat{n}_{\zeta}$
can be approximately factored into a product of three cartesian rotations,
i.e., 
\begin{equation}
\exp{( i \zeta \hat{J} \cdot \hat{n}_{\zeta})}  \simeq  
		\exp{(i \zeta  J_x \sin{\theta}\cos{\phi})} 
		\exp{(i \zeta  J_y \sin{\theta}\sin{\phi})} 
		\exp{(i \zeta  J_z \cos{\theta})}.
\end{equation}
\label{app:rotation_expand}
Then the corresponding one-dimensional rotation around, e.g., the $\hat{x}$ 
axis, is made with the small angle $\zeta  \sin{\theta}\cos{\phi}$.

\subsection{Commutation Relations} 
Expanding the standard commutator
\begin{eqnarray}
\left[\frac{\partial}{\partial x}, f(x) \right] & = & \frac{\partial f(x)}{\partial x},  
\end{eqnarray}
we arrive at 
\begin{eqnarray}
 f(x) \frac{\partial}{\partial x} 
& = & \frac{\partial}{\partial x} f(x) -\frac{\partial f(x)}{\partial x} \\
& = & \frac{i}{\hbar} \hat P_x f(x) -\frac{\partial f(x)}{\partial x}. 
\label{app:partial}
\end{eqnarray}
Generalizing the one-dimensional function $f(x)$ to
the three-dimensional
$\vec{F}(\vec{R}) = f_x \vec{i} + f_y \vec{j} + f_z \vec{k}$,
and evaluating Eq.~(\ref{app:partial}) for each degree of freedom,
leads then to the desired relation
\begin{equation}
\vec{F} \cdot \vec{\nabla} = \frac {i} {\hbar} \hat{P} \cdot \vec{F}
                                -\vec{\nabla} \cdot \vec{F}.
\end{equation}


\subsection{Computational Scheme}

Start with an initial ensemble of walkers (e.g. distributed
according to the trial wavefunction). Compute the local
energy (also used as the starting reference energy) for the initial 
ensemble and then propagate the ensemble in imaginary time by 
repeating the following steps for each walker (configuration).

\begin{enumerate}

\item\label{trans}  Move each host and impurity atom according
to equation~(\ref{eq:trans_move}). This move involves a Gaussian
random number and a quantum force.

\item\label{transmet} Each move is then accepted or rejected with acceptance
probability given by $min(1,W(\vec{Q}',\vec{Q}))$ where
$$
W(\vec{Q}',\vec{Q}) = \frac{|\Psi_T(\vec{Q}')|^2}{|\Psi_T(\vec{Q})|^2} 
\frac{G(\vec{Q}' \rightarrow \vec{Q}, \Delta \tau)}{G(\vec{Q} \rightarrow \vec{Q}', \Delta \tau)}
$$

This involves having to recompute the quantum force after the particle
has been moved.

\item\label{rot}  Rotate the impurity atom according to equation~(\ref{eq:rot_move})

\item\label{rotmet}  Again, this move is accepted or rejected using the
above criterion.

\item\label{dt}  Determine effective time step, $\Delta \tau_{eff}$. See Ref.~\cite{reynolds82}. One has to average the effective time step over
the various kinds of moves since in general, each kind of move has a 
different diffusion constant.

\item\label{update}  Compute the new local energy, $E_L(\vec{Q}')$
(equation~(\ref{eq:e_local}))

\item\label{branch}  Compute branching factor, $M$:

$$
M = \rm{int} \left(\exp \left[\left(E_R - \frac{E_L(\vec{Q}) + E_L(\vec{Q}')}{2}\right)\Delta \tau_{eff} \right] + \Delta \right) 
$$
where $\Delta$ is a random number that is uniformly distributed over (0,1).
If $M=0$, this walker is destroyed, otherwise $M-1$ duplicates of the
configuration are added to the ensemble.

\item\label{e_ref} To maintain a stable population size (N($\tau$)), update 
the reference energy according to
$$
E_R(\tau + \Delta \tau) = E_R(\tau) + \frac{\alpha}{\Delta \tau}
\log \frac{N^{*}}{N(\tau+\Delta \tau)}
$$
where the population control parameter, $\alpha$ is chosen to be as small as
possible (to avoid biasing the results) but large enough 
that the population size stays within acceptable limits. $N^*$ is the
desired population size and is often chosen to be the starting population
$N(0)$ or  $N(\tau)$.  
\end{enumerate}

Of course, many variants of the above procedure will also work.  In the
HCN studies, for example, all of the translation moves and rotations are
performed simultaneously and then accepted or rejected as a single move.
When studying excited states using the fixed-node approximation, one also 
adds the additional constraint that moves crossing the nodal surface are 
automatically rejected.

The statistical error is estimated using block averaging, where the propagation
is split into $N_b$ blocks of $N_s$ steps where $N_s$ is 
longer than the correlation length\cite{hammond94}.
In the literature, various methods are used to 
compute averages and errors from this blocked data:
\begin{enumerate}
\item Take 1 data point per block, and compute the average and
standard error from these $N_b$ data points
\item Same as above but use the {\em block average} values for the
$N_b$ data points
\item Use all the data points in computing the mean. The standard error
for each block, 
$\sigma_b$, is computed from the $N_s$ values within each block.  These 
are averaged to determine the reported error: 
$$
\sigma = \frac{\overline{\sigma_b}}{\sqrt{N_b - 1}}
$$
\end{enumerate}

In our examples, we report the largest of the above estimates (usually
close in value).

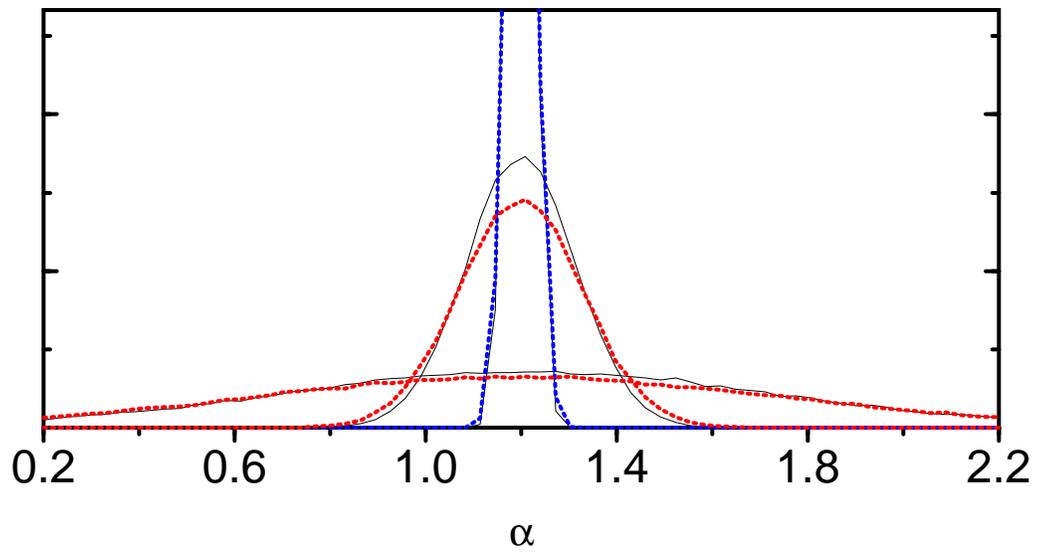
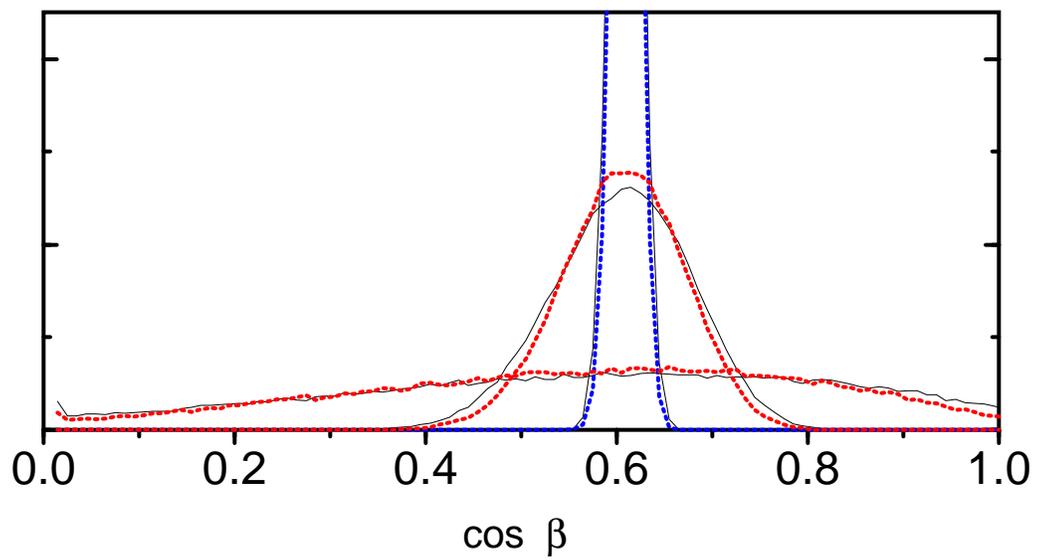
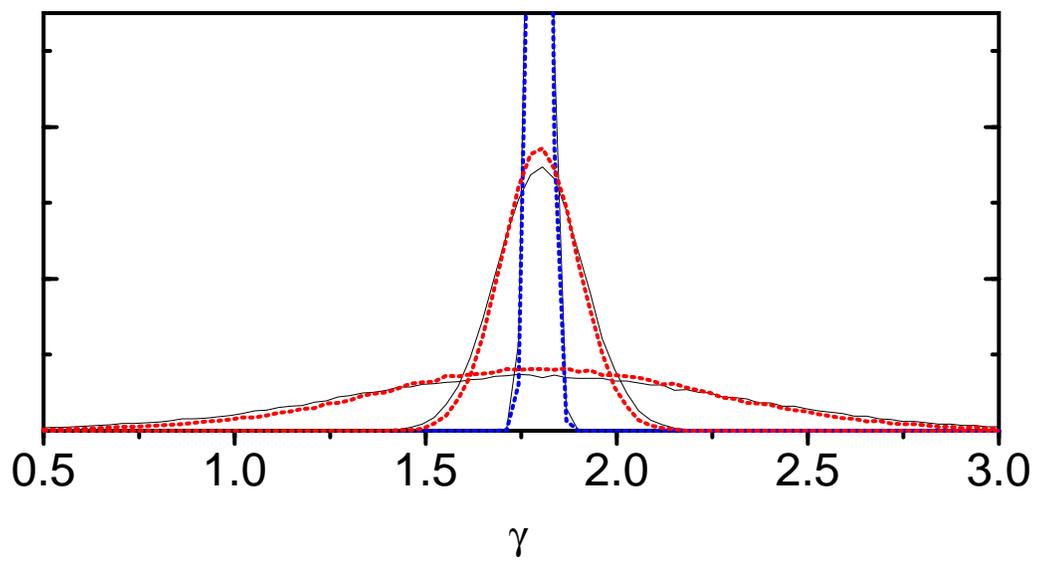
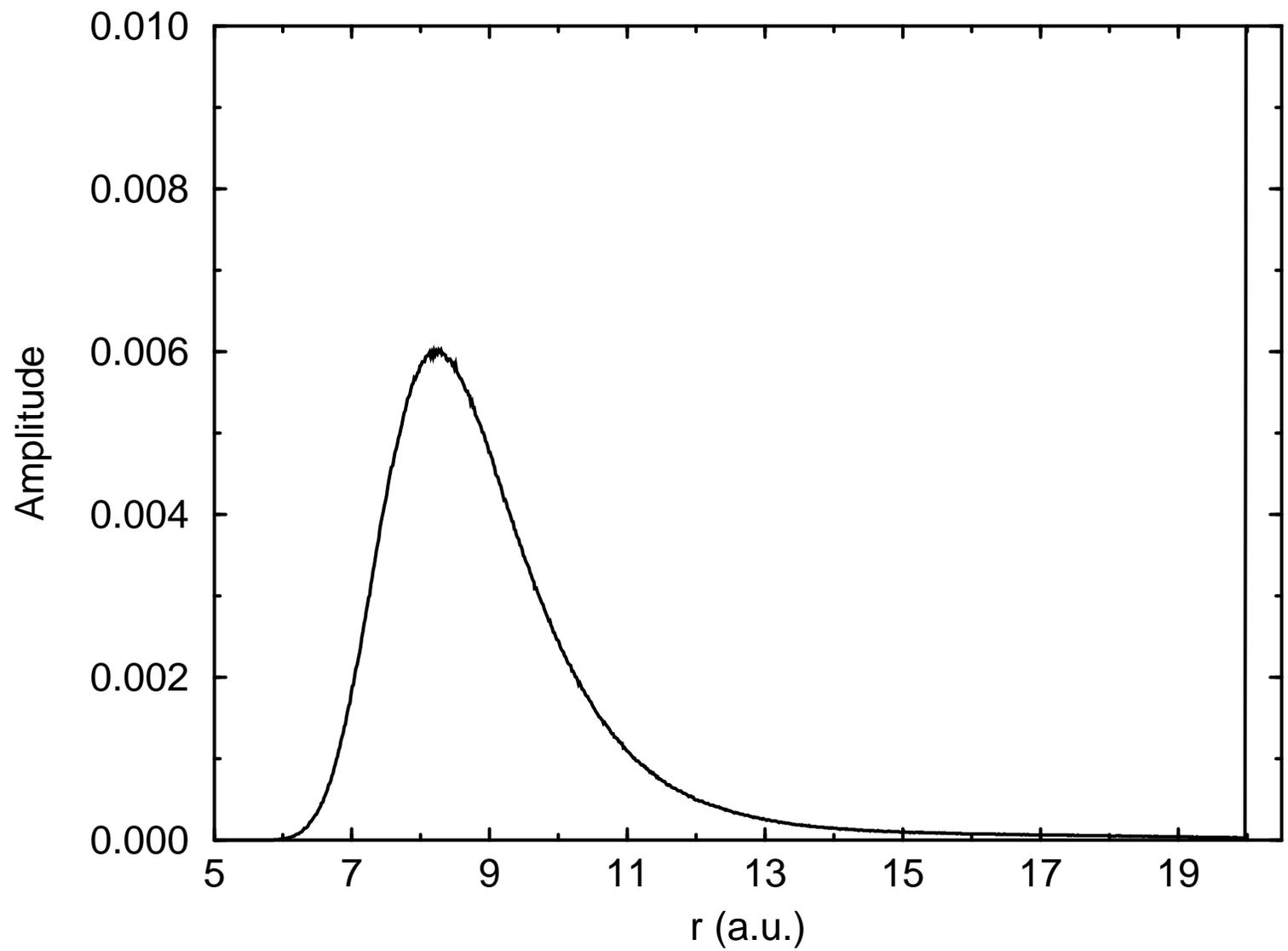
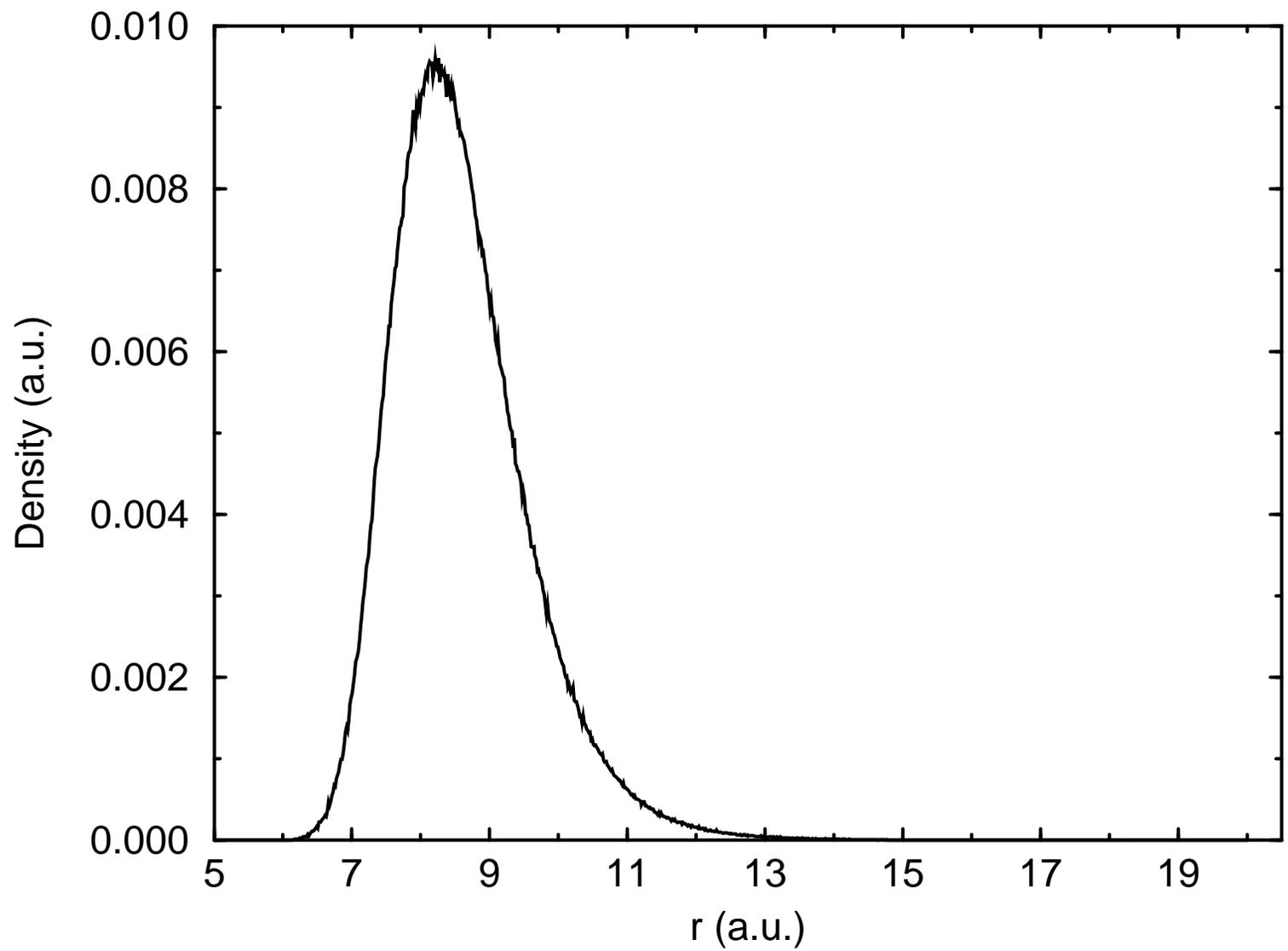
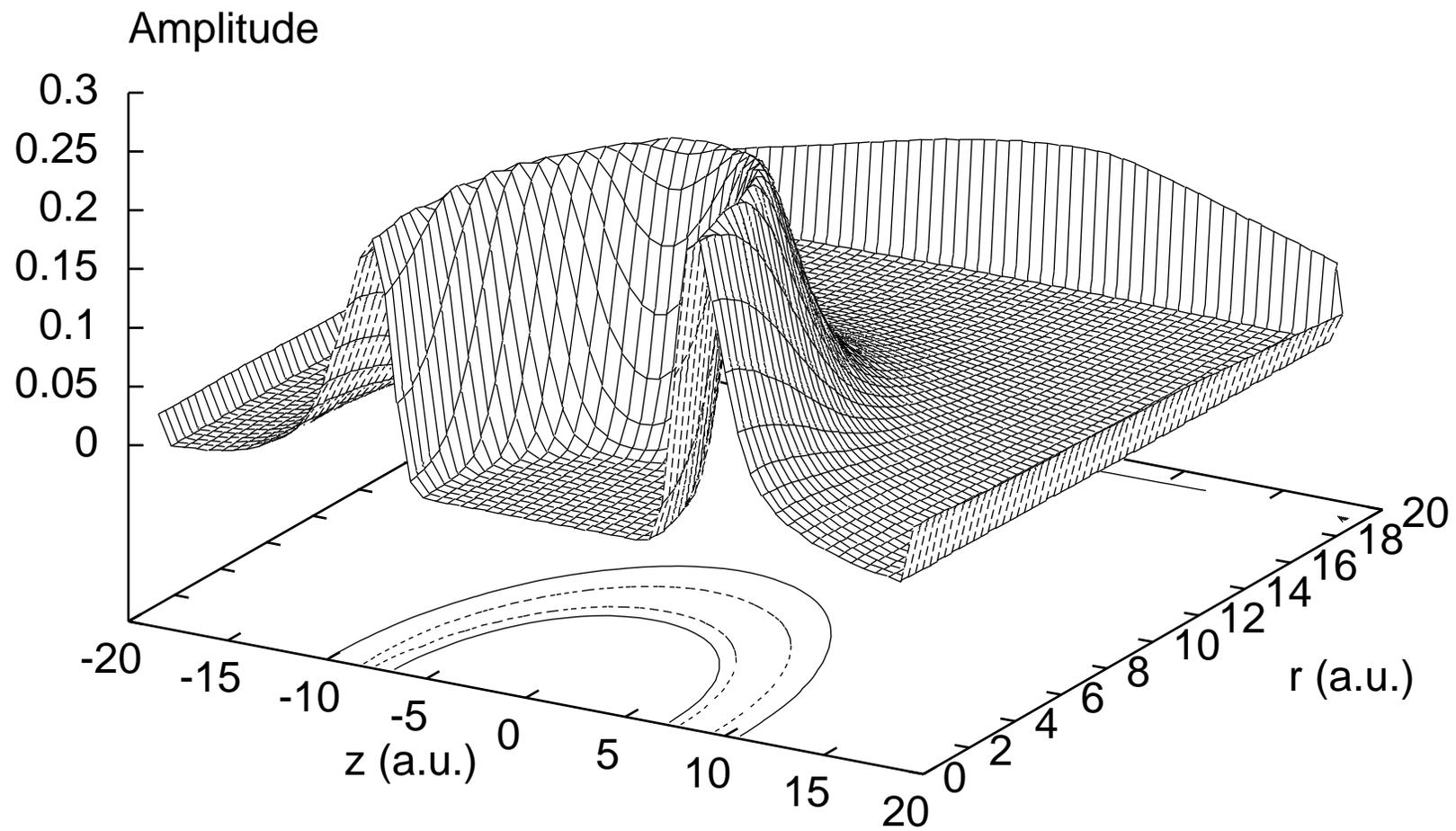
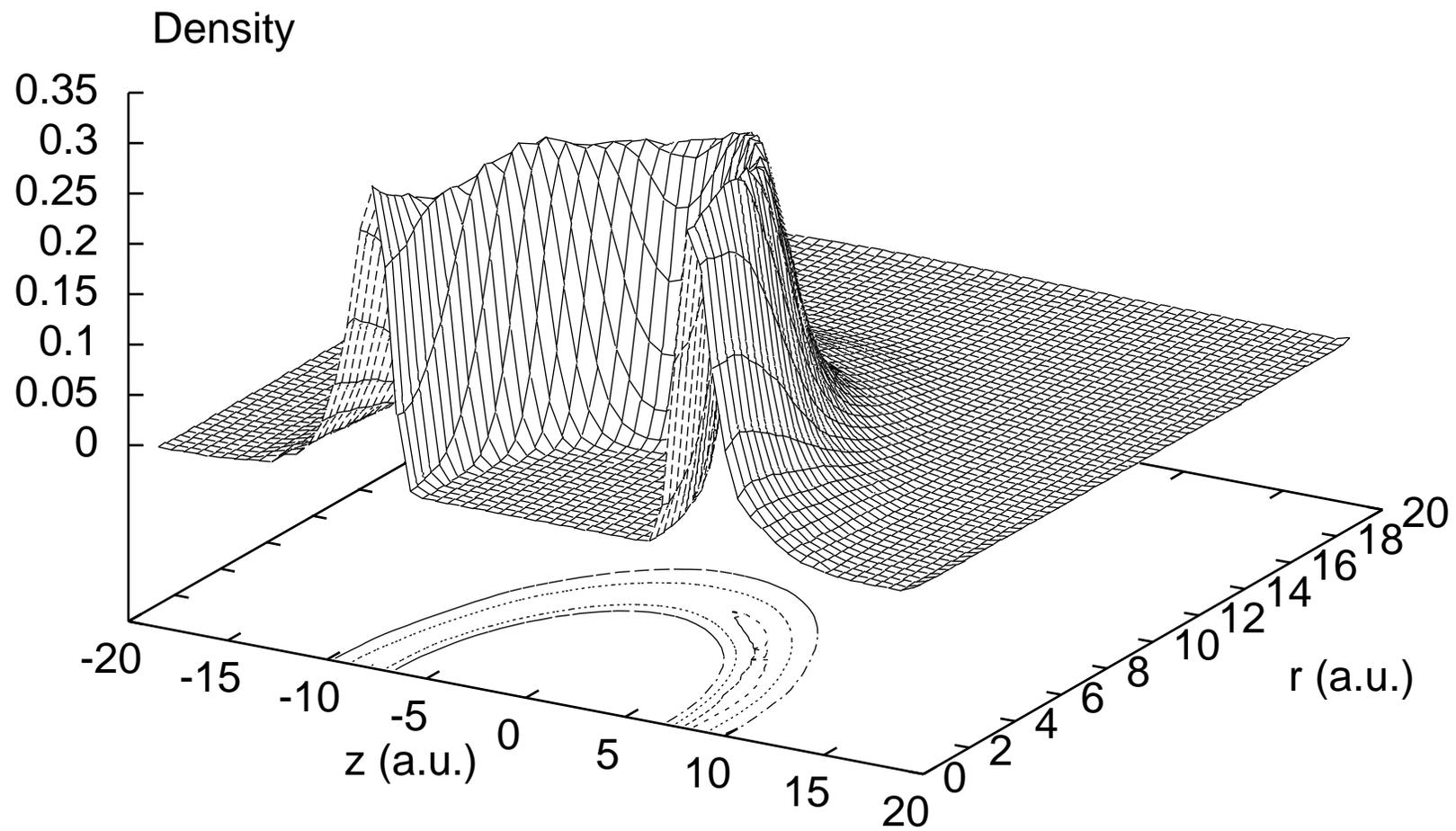
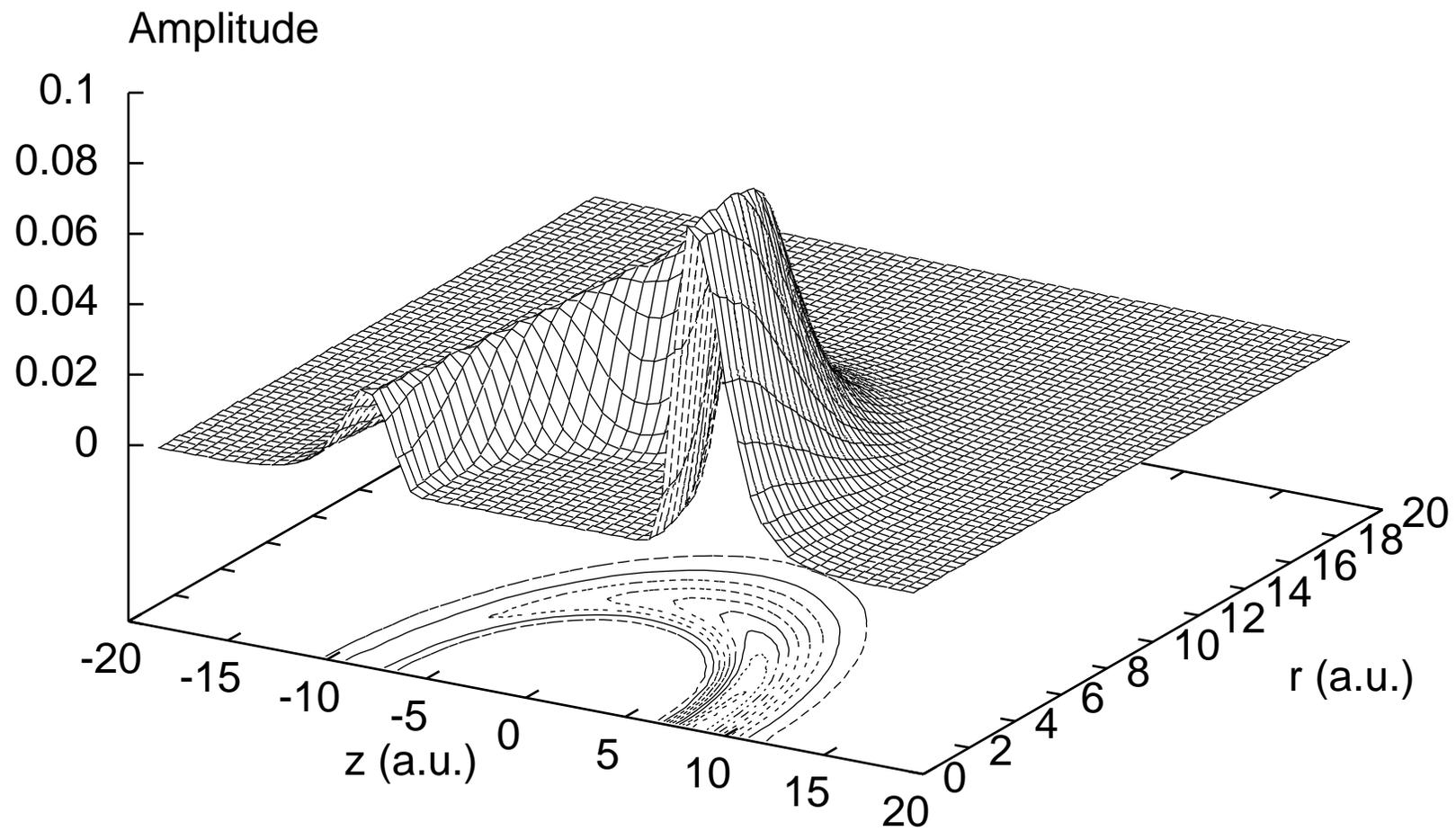
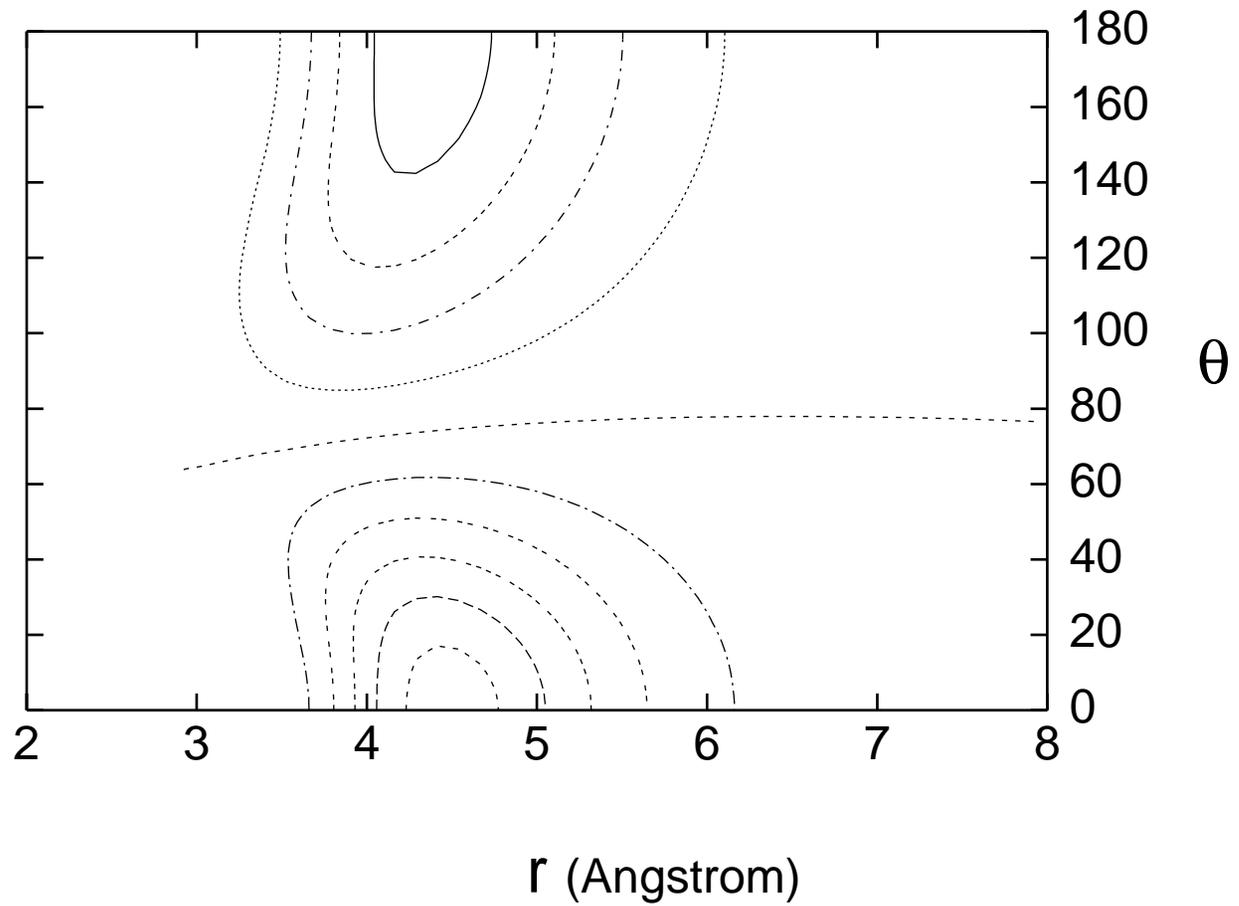
\end{document}